**Rotating, cylindrical, acoustic invisibility cloak: solution using perturbation method**

Levi T. Kaganowich, Deepak C. Akiwate, Trevor J. Cox, Olga Umnova

**Abstract**

Transformation Acoustics emerged in the mid-2000s, initiating a new paradigm of metamaterial designs. One of the most compelling designs, the invisibility cloak, holds promise for stealth and noise reduction applications in aviation. However, adapting this design to meet the demands of realistic conditions has proven challenging. This work focusses on the design of a stationary 2D cylindrical cloak and its performance whilst rotating, a result not yet reported in the literature. The study utilises linearised equations of motion with convective terms. A wave equation is derived, and corresponding solution and scattering coefficient are derived using a perturbation method and exact numerical solution. These results are used to evaluate the performance of the rotating cloak. Results show that rotation causes a reduction in cloaking performance with greater scattering observed for increasing rotational speeds, with a reasonable agreement (within 5%) between methods over the range of applicability. The perturbation method offers a fast, computationally inexpensive means of evaluating a rotating, graded anisotropic fluid.

**Nomenclature**

$c$ – wave speed

$f$ – frequency

$i$ – imaginary unit

$k$ – wavenumber

$m$ – harmonic number

$p$ – pressure

$r$ – radius

$\boldsymbol{v}$ – acoustic velocity

$\boldsymbol{u}$ – acoustic displacement

$\boldsymbol{v_0}$ – fluid background velocity

$\theta$ – polar angle

$\lambda$ – bulk modulus

$\rho$ – density (scalar)

$\bar{\bar{\rho}}$ – density (tensor)

$\bar{\rho}$ – normalised density (tensor), $\bar{\bar{\rho}}/\rho$

$\omega$ – angular frequency

$\Omega$ – rotational speed

## 1.1 Introduction

Environmental noise pollution is recognised by the World Health Organisation as a significant contributing factor to adverse health outcomes in our society (World Health Organisation, 2022). In particular, aviation is a major source of noise pollution, and increasingly strict noise reduction targets have been set in a drive to make the sector more sustainable. For example, the Advisory Council for Aviation Research and Innovation in Europe (ACARE) has an ambitious 15dB (65%) reduction target in perceived noise relative to the 2000 baseline by 2050 (ACARE, 2024). New technologies and analytical models need to be developed to meet this goal (Knobloch et al. 2022), especially drawing upon the potential of metamaterials (Palma et al., 2018). The following study aims to contribute towards this effort by understanding the use of a transformation acoustics method under a realistic condition, with an aeroacoustic application in mind.

Transformation acoustics invisibility cloaking is a technique that finds the acoustic material parameters needed to control propagation such that a target region is rendered acoustically invisible, based on the results of Pendry et al. (2006), in optics. By exploiting the form invariance of the acoustic equations of motion, a coordinate transformation is interpreted as an input, generating physical material parameters that achieve the desired propagation (Cummer et al., 2008; Chen and Chan 2007; 2010). The ideal cloak design causes waves to be smoothly guided around the target region so that scattering from this region is completely cancelled. The resulting design parameters are usually acoustically inhomogeneous or anisotropic and the material acts as an acoustic fluid (Cummer, 2013). Whilst various cloak designs have been studied since the result was established, this work will focus on the 2D cylindrical cloak, the benchmark cloak design in the literature.

The transformation acoustics technique is valid in static conditions. Movement such as rotation or flow introduces convective terms for which the form invariance across a spatial transformation breaks down (García-Meca et al., 2013). Many efforts have been made to overcome this barrier, with a noted focus on the potential aeroacoustic applications of the 2D cylindrical cloak. For example, the simple modification of parameters to account for a mean flow (Zhong & Huang, 2013; Huang et al., 2014;

Iemma 2016), as well as the use of differential geometry to establish a transformation that is both temporal and spatial (García-Meca et al., 2014a; 2014b; Iemma & Palma 2017; 2018; 2020; Colombo et al., 2024). To date, no work has modelled the effect of rotation on the acoustic 2D cylindrical cloak. However, a more recent work has modelled the effect of rotation on a 2D cylindrical electromagnetic cloak (Hasanpour Tadi & Shokri, 2024). Also relevant is a study which has established an acoustic cloaking effect through contrarotating homogeneous fluid cylinders (Farhat et al., 2020).

The effects of rotation on an acoustic fluid have been studied under idealised conditions showing that rotation causes scattering. For a homogeneous fluid, it is possible to derive an exact solution for a rotating 2D cylinder, as shown by Censor and Aboudi (1971). They demonstrate that increasing rotational speed causes more scattering. Their method does not account for viscous effects or turbulence, limiting its validity to slower rotational speeds. The present work adapts their method to model the wave propagation in an inhomogeneous, anisotropic rotating fluid deriving semi-analytical and numerical solutions to the resulting wave equation, to evaluate the performance of a rotating invisibility cloak.

### 2.1 Methods

The main aim of this study is to evaluate the performance of a rotating cloak. The main propagation feature of an invisibility cloak is scattering cancellation, therefore, the outgoing scattered wave from a rotating cloak will be the main focal point of the study. The set up of the study is described below in Fig. 1. An incident plane wave from the left meets the cloak $(a < r \leq b)$, rotating at constant speed, $\Omega$. The cloaked region is $r < a$ and the background fluid is $r > b$.

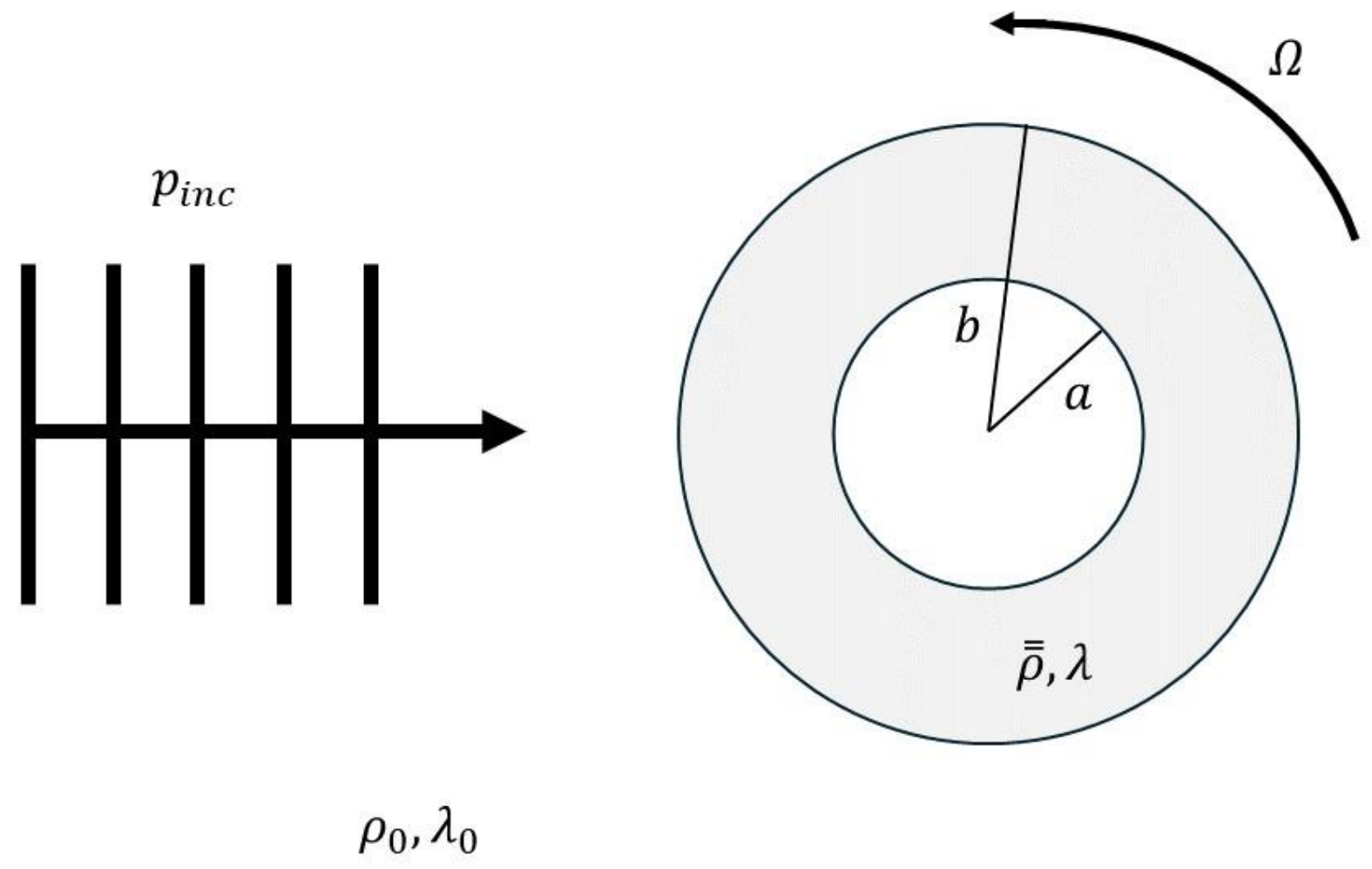

Fig. 1. Diagram of rotating cloak with incident plane wave

To perform the study, the method of transformation acoustics is used to derive the parameters of an ideal 2D cylindrical cloak, as well as another case: the near cloak. Next, the equations of motion for a rotating fluid are applied to a rotating cloak medium of unspecified parameters, leading to a second order wave equation in the frequency domain. Since the equation appears not to have an analytical solution, a perturbation expansion and matching solution at the cloak outer radius are proposed. This solution will be evaluated for the ideal cloak parameters in a set of test conditions and compared to a direct numerical solution developed from the equation.

**2.1.1 Transformation acoustics cloak**

The method of transformation acoustics feeds the equations of linear acoustics through a coordinate transformation chosen to hide a region of space. This hinges on the use of the Jacobian matrix for orthogonal coordinate transformations. The transformed parameters – bulk modulus and density – can then be interpreted as material parameters within the original coordinate system (Cummer, 2013).

Recall the equations of linear acoustics, noting assumptions of an adiabatic, isentropic and homentropic fluid, with sound pressure levels below 140dB. Throughout this study, waves are assumed to be time harmonic with $e^{-i\omega t}$ dependence.

$$\nabla p = i\omega\rho_0\boldsymbol{v} \tag{1}$$

$$i\omega p = \lambda_0\nabla.\boldsymbol{v} \tag{2}$$

Applying a coordinate transformation to the equations of linear acoustics leads to the following equations, where $A$ is the Jacobian matrix of transformation:

$$\nabla p = i\omega[\det(A)\,(A^T)^{-1}\rho_0(A^{-1})]\boldsymbol{v} \tag{3}$$

$$i\omega p = [\lambda_0 \det(A)]\nabla.\boldsymbol{v} \tag{4}$$

Noticing that, fundamentally, the form of the equations hasn't changed. The terms in the square brackets are the 'instruction' for achieving the transformation and according to transformation acoustics, can be interpreted as new material parameters.

$$\bar{\bar{\rho}} = \det(A)(A^T)^{-1}\rho_0(A^{-1}) \tag{5}$$

$$\lambda = \lambda_0\det(A) \tag{6}$$

Where $\rho_0$ refers to the density in the fluid outside of the cloak and is scalar, $\bar{\bar{\rho}}$ refers to the density inside the cloak and is a tensor, $\lambda_0$ refers to the bulk modulus in the fluid outside of the cloak and is scalar, and $\lambda$ refers to the bulk modulus inside the cloak. Both new material parameters are graded: they are a function of radius.

These new parameters in the physical region are no longer homogeneous; the density is anisotropic and inhomogeneous, and the bulk modulus is homogeneous. Note that anisotropic density does not imply a different weight in orthogonal directions but a different inertial response.

The coordinate transformation for a 2D cylindrical cloak described by Cummer (2013) is as follows:

$$r = \frac{(b-a)}{b}r' + a \tag{7}$$

$$\theta = \theta' \tag{8}$$

Valid for $0 < r' \leq b$ and $a < r \leq b$. (Cummer, 2013, p.209)

Applying this transformation leads to the following material parameters for density (expressed as an inverse for convenience) and bulk modulus:

$$\bar{\bar{\rho}}^{-1} = \rho_0^{-1}\begin{pmatrix} \frac{r-a}{r} & 0 \\ 0 & \frac{r}{r-a} \end{pmatrix} \tag{9}$$

$$\lambda = \lambda_0 \frac{(b-a)^2}{b^2}\frac{r}{r-a} \tag{10}$$

Inspection of these parameters reveals a singularity at the cloak inner radius $r = a$. This feature may introduce difficulties, especially when seeking a direct numerical solution. The problematic nature of the material parameters required close to the inner radius of the ideal cloak are well acknowledged in the literature (Ruan et al., 2007; Kohn et al., 2010; Torrent & Sánchez-Dehesa, 2008; Sánchez-Dehesa & Torrent, 2013). Norris (2008) points out that the ideal design requires the cloak to be infinitely massive at the inner radius and suggests alternative formulations to avoid the singularity such as a near cloak and elastic Pentamode Cloak, see also Gokhale et al., (2012) for more details on alternative transformations.

A study of the effects of rotation on the Pentamode invisibility cloak are beyond the scope of this work. However, the near cloak described by Norris (2008) makes a minor change to transformation (7). Therefore, it will be useful to include the near cloak within this study to avoid problems with the singularity in the numerical solution. By introducing a small parameter $\delta$, transformation (7) now becomes:

$$r = \frac{r'(b-a) - b(\delta - a)}{b-\delta} \tag{11}$$

$$\theta = \theta' \tag{12}$$

Valid for $\delta < r' \leq b$ and $a < r \leq b$

This generates the following cloak parameters:

$$\bar{\bar{\rho}}^{-1} = \rho_0^{-1}\begin{pmatrix} \frac{r(b-\delta) + b(\delta - a)}{r(b-\delta)} & 0 \\ 0 & \frac{r(b-\delta)}{r(b-\delta) + b(\delta - a)} \end{pmatrix} \tag{13}$$

$$\lambda = \lambda_0 \frac{(b-a)}{(b-\delta)}\frac{r(b-a)}{\big(r(b-\delta) + b(\delta - a)\big)} \tag{14}$$

The near cloak is almost identical to the ideal cloak except for a small inclusion of radius $\delta$ in the original coordinate system. There is a divergence in behaviour very close to the inner radius. This is visually demonstrated in the Fig. 2. below where ideal and near cloak parameters are plotted to show their close similarity, for a typical cloak size in the results section.

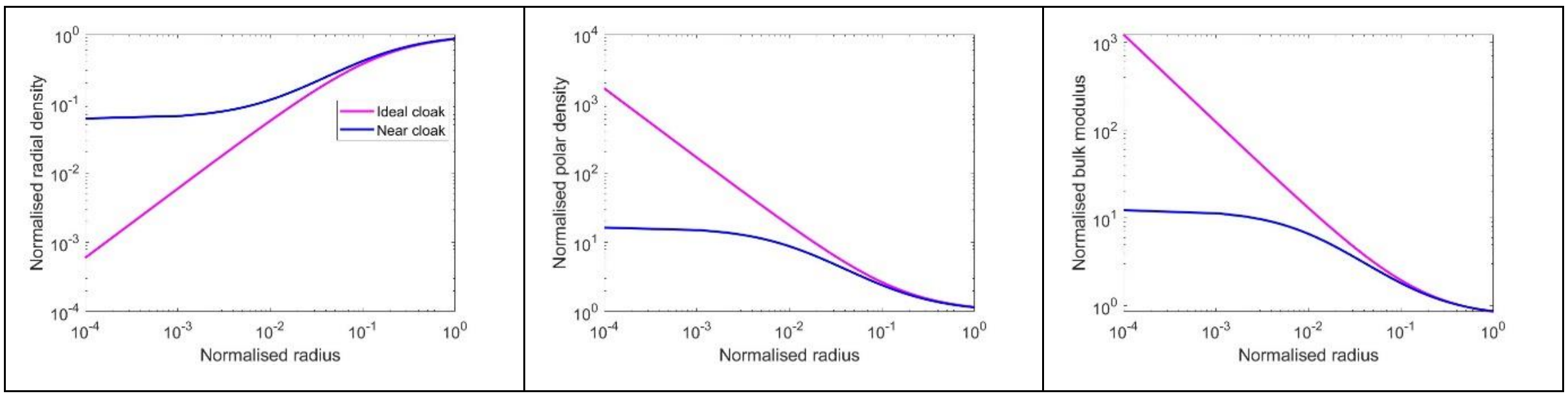

Fig. 2. Graph plotting normalised radius versus normalised ideal and near cloak parameters

### 2.1.2 Equations of motion for a rotating cloak

Having described the cloak parameters, the next step is to develop a set of equations of motion for a rotating cloak. Whilst the specific parameters for an ideal cloak were derived in the previous section, the equations derived in this section are for unspecified cloak parameters. This will allow a general formulation of the governing wave equation as well as the opportunity to test modified cloak parameters as part of further research. It is assumed that the bulk modulus parameter is a scalar and density parameter is a tensor.

Adopting the method of Censor and Aboudi (1971) and Farhat et al., (2020), the following mass and momentum equations are defined for a cloak with uniform rotation, $\boldsymbol{v_0} = \Omega r\boldsymbol{\hat{\theta}}$, and acoustical velocity, $\boldsymbol{v}$:

$$(\partial/\partial t + \boldsymbol{v}_0 \cdot \boldsymbol{\nabla})\boldsymbol{v} + (\boldsymbol{v} \cdot \boldsymbol{\nabla})\boldsymbol{v}_0 = -\bar{\bar{\rho}}^{-1}\boldsymbol{\nabla} p \tag{15}$$

$$(\partial/\partial t + \boldsymbol{v}_0 \cdot \boldsymbol{\nabla})p + \lambda \boldsymbol{\nabla} \cdot \boldsymbol{v} = 0 \tag{16}$$

Combining these equations and assuming that wave propagation is periodic in the polar direction with dependence $e^{im\theta}$ such that $\partial/\partial\theta \equiv im$, it is possible to derive the following general wave equation for a rotating, anisotropic, graded, cylindrical fluid region – previously not described in the literature. This equation can be used to model a rotating cloak of unspecified parameters. Equation (17):

$$\frac{\partial^2 p}{\partial r^2}+\frac{\partial p}{\partial r}\left(\frac{1}{r}+\frac{1}{\bar{\bar{\rho}}_r^{-1}}\frac{\partial \bar{\bar{\rho}}_r^{-1}}{\partial r}+\frac{2\Omega im}{Mr}\left(\frac{\bar{\bar{\rho}}_\theta^{-1}}{\bar{\bar{\rho}}_r^{-1}}-1\right)\right)+p\left(\frac{2\Omega im}{Mr\bar{\bar{\rho}}_r^{-1}}\frac{\partial \bar{\bar{\rho}}_\theta^{-1}}{\partial r}-\frac{m^2}{r^2}\frac{\bar{\bar{\rho}}_\theta^{-1}}{\bar{\bar{\rho}}_r^{-1}}-\frac{(M^2+4\Omega^2)}{\bar{\bar{\rho}}_r^{-1}\lambda}\right)=0$$

Where $M=-i\omega+im\Omega$

Unlike the result for a rotating homogenous, isotropic fluid cylinder shown by Censor and Aboudi (1971), this wave equation no longer holds the form of the homogenous Bessel's equation and does not appear to have an analytical solution. The following subsections will attempt to find a solution to equation (17), in order to evaluate the performance of a rotating cloak.

### 2.1.3 Perturbation expansion and matching solution

The first method employed to solve equation (17) is a perturbation expansion and matching solution. This involves stating the boundary conditions outside of the cloak, and combining them into one quotient that can be identically matched to a solution of a Riccati type equation developed from (17). Finally, a perturbation expansion is developed, applied to the equation and solved to first order. Therefore, this method will only be valid under certain conditions to be specified. This approach is taken since the analytical solution to the pressure field inside the rotating cloak is unknown.

First, state the boundary conditions to be solved: continuity of pressure and normal displacement. These are combined into one quotient, and evaluated at the cloak outer radius, $r=b$:

$$\frac{u_r(b)}{p(b)}=\frac{k\big(J_m'(kb)+A_mH_m'(kb)\big)}{\omega^2\rho_0\big(J_m(kb)+A_mH_m(kb)\big)} \tag{18}$$

Continuity of normal displacement is chosen as a boundary condition, following the rationale of Morse and Ingard (1968, p. 711). Now state the same boundary conditions inside the cloak, noting that displacement now includes the effect of rotation (Censor & Aboudi, 1971, p. 440):

$$\frac{u_r}{p} = \frac{(2\Omega^2 - M^2)\frac{\bar{\bar{\rho}}_r^{-1}}{p}\frac{\partial p}{\partial r} - 3\Omega M \bar{\bar{\rho}}_\theta^{-1}\frac{im}{r}}{(4\Omega^2 + M^2)(M^2 + \Omega^2)} \tag{19}$$

Note that $u_r/p$ inside the cloak depends on $\partial p/\partial r$, which is unknown. To deal with this, a Riccati type equation is developed for $u_r/p$, using (17) and (19). This leads to the following equation (20), rewriting $u_r/p$ as $x$, where $'$ indicates $\partial/\partial r$ unless otherwise stated:

$$x_m' + x_m^2 \frac{1}{\bar{\bar{\rho}}_r^{-1}} \frac{(M^2 + \Omega^2)(M^2 + 4\Omega^2)}{2\Omega^2 - M^2} + x_m \frac{1}{r}\left(1 + \frac{2im\Omega}{M}\left(2\frac{\bar{\bar{\rho}}_\theta^{-1}}{\bar{\bar{\rho}}_r^{-1}}\frac{M^2 + \Omega^2}{2\Omega^2 - M^2} - 1\right)\right)$$
$$+ \frac{\bar{\bar{\rho}}_\theta^{-1}}{M^2 + \Omega^2}\left(\frac{1}{r}\frac{im\Omega}{M}\frac{\left(\bar{\bar{\rho}}_\theta^{-1}\right)'}{\bar{\bar{\rho}}_\theta^{-1}} - \frac{1}{r^2}\frac{\bar{\bar{\rho}}_\theta^{-1}}{\bar{\bar{\rho}}_r^{-1}}\frac{3m^2\Omega^2}{2\Omega^2 - M^2} + \frac{m^2}{r^2} - \frac{2\Omega^2 - m^2}{\lambda\bar{\bar{\rho}}_\theta^{-1}}\right) = 0$$

This equation will be solved at $r = b$, and solution can be matched to (18). This allows for the scattering coefficient $A_m$ and outgoing wave found, $A_m H_m(kr)$ to be found, to evaluate the cloaking performance.

From here, it will be desirable to express equation (20) in a normalised, dimensionless form, as follows, equation (21):

$$X_m' + X_m^2 \frac{q}{\bar{\rho}_r^{-1}} \frac{(\boldsymbol{M}^2 + \boldsymbol{\Omega}^2)(\boldsymbol{M}^2 + 4\boldsymbol{\Omega}^2)}{(2\boldsymbol{\Omega}^2 - \boldsymbol{M}^2)} + X_m \frac{1}{R_n + A}\left(1 + \frac{2im\boldsymbol{\Omega}}{\boldsymbol{M}}\left(2\frac{\bar{\rho}_\theta^{-1}}{\bar{\rho}_r^{-1}}\frac{\boldsymbol{M}^2 + \boldsymbol{\Omega}^2}{2\boldsymbol{\Omega}^2 - \boldsymbol{M}^2} - 1\right)\right)$$
$$+ \frac{1}{q}\frac{\bar{\rho}_\theta^{-1}}{\boldsymbol{M}^2 + \boldsymbol{\Omega}^2}\left(\frac{1}{R_n + A}\frac{im\boldsymbol{\Omega}}{\boldsymbol{M}}\frac{\left(\bar{\rho}_\theta^{-1}\right)'}{\bar{\rho}_\theta^{-1}} - \frac{1}{(R_n + A)^2}\frac{\bar{\rho}_\theta^{-1}}{\bar{\rho}_r^{-1}}\frac{3m^2\boldsymbol{\Omega}^2}{2\boldsymbol{\Omega}^2 - \boldsymbol{M}^2} + \frac{m^2}{(R_n + A)^2}\right.$$
$$\left. - q^2\frac{2\boldsymbol{\Omega}^2 - \boldsymbol{M}^2}{\lambda\bar{\rho}_\theta^{-1}}\right) = 0$$

Which relies on the following normalised and dimensionless quantities:

$$\rho_0 c\omega x = X \tag{22}$$

$$\boldsymbol{\Omega} = \frac{\Omega}{\omega}, \qquad \boldsymbol{M} = \frac{M}{\omega} \tag{23}$$

$$q = \frac{\omega}{c}(b - a) = k(b - a) \tag{24}$$

For the ideal cloak, the material parameters will be:

$$\bar{\rho}_r^{-1} = \frac{R_n}{R_n + A}, \qquad \bar{\rho}_\theta^{-1} = \frac{R_n + A}{R_n}, \qquad \lambda = \frac{1}{(1+A)^2}\frac{R_n + A}{R_n} \tag{25}$$

Where $R_n = \frac{R}{b-a}$, $R = r - a$, $A = \frac{a}{b-a}$, subscript $n$ indicates normalisation.

The boundary conditions (18) now become:

$$\frac{J_m'(kb) + A_m H_m'(kb)}{J_m(kb) + A_m H_m(kb)} = X_m(R_n = 1) \tag{27}$$

To solve (21), define a perturbation term $\epsilon_m$ and first order expansion in terms of $X$:

$$\epsilon_m = m\left(\frac{\Omega}{\omega}\right) \tag{28}$$

$$X_m = X_m^{(0)} + X_m^{(1)}\epsilon_m, \qquad m \in \mathbb{Z}, m \neq 0 \tag{29}$$

Where terms $X_m^{(2)}\epsilon_m^2$ and greater are disregarded, which necessitates that $\Omega$ remains small enough compared to $\omega$ such that $|\epsilon_m| \ll 1$, with care also being taken with the number of harmonic terms used in the calculation. Therefore, lower frequencies and smaller rotational speeds will be chosen to ensure solutions remain valid.

To account for $X_0$, (when harmonic $m = 0$), it is necessary to consider $\left(\frac{\Omega}{\omega}\right)^2$ which is the leading order term in this case. An analytical solution exists for $X_0$, and this will be included in the expansion. Whilst the overall contribution of this solution will be comparatively small, and usually negligible, it is included for completeness.

Substituting the perturbation term (28) into equation (21), leads to the following equation (30):

$$X_m' + \frac{X_m^2 q}{\bar{\rho}_r^{-1}}(1 - 2\epsilon_m) + \frac{X_m}{R_n + A}\left(1 + 2\epsilon_m\left(\frac{2\bar{\rho}_\theta^{-1}}{\bar{\rho}_r^{-1}} + 1\right)\right) - \frac{1}{q}\bar{\rho}_\theta^{-1}\left(-\epsilon_m\frac{\left(\bar{\rho}_\theta^{-1}\right)'}{(R_n + A)\bar{\rho}_\theta^{-1}} + \frac{m^2}{(R_n + A)^2}(1 + 2\epsilon_m) - \frac{q^2}{\lambda\bar{\rho}_\theta^{-1}}\right) = 0$$

Substitution of expansion (29) leads to the following zeroth and first order equations for unspecified cloak parameters:

$$\left(X_m^{(0)}\right)' + \frac{\left(X_m^{(0)}\right)^2 q}{\bar{\rho}_r^{-1}} + \frac{X_m^{(0)}}{R_n + A} - \frac{1}{q}\bar{\rho}_\theta^{-1}\left(\frac{m^2}{(R_n + A)^2} - \frac{q^2}{\lambda\bar{\rho}_\theta^{-1}}\right) = 0 \qquad (31)$$

First order equation (32):

$$\left(X_m^{(1)}\right)' + X_m^{(1)}\left(\frac{q}{\bar{\rho}_r^{-1}} 2X_m^{(0)} + \frac{1}{R_n + A}\right) + \left(-2\frac{q}{\bar{\rho}_r^{-1}}\left(X_m^{(0)}\right)^2\right) + \frac{1}{R_n + A}\left(2X_m^{(0)}\left(\frac{2\bar{\rho}_\theta^{-1}}{\bar{\rho}_r^{-1}} + 1\right)\right)$$
$$+ \frac{1}{q}\bar{\rho}_\theta^{-1}\left(+\frac{\left(\bar{\rho}_\theta^{-1}\right)'}{(R_n + A)\bar{\rho}_\theta^{-1}} - \frac{m^2 2}{(R_n + A)^2}\right) = 0$$

For the ideal cloak parameters (25) the zeroth order equation (31) becomes:

$$\left(X_m^{(0)}\right)' + \left(X_m^{(0)}\right)^2 \frac{q(R_n + A)}{R_n} + \frac{X_m^{(0)}}{R_n + A} - \frac{1}{q(R_n + A)}\left(\frac{m^2}{R_n} - \frac{q^2(1 + A)^2 R_n}{1}\right) = 0 \qquad (33)$$

The solution to this equation is:

$$X_m^{(0)} = \frac{J_m'(z)}{J_m(z)} \frac{z(1 + A)}{z + A'} \qquad (34)$$

Where $q' = (1 + A)q$, $A' = Aq'$, and $z = q'R_n$

In the special case of $X_0$, where $m = 0$, an exact solution to (21) is available, where $\boldsymbol{\Omega}^2$ is the leading order term:

$$X_0 = \frac{(2\boldsymbol{\Omega}^2 + 1)}{(1 - \boldsymbol{\Omega}^2)} \frac{R_n}{R_n + A}(1 + A)\frac{J_0'\left(kb\sqrt{1 - 4\boldsymbol{\Omega}^2}R_n\right)}{J_0\left(kb\sqrt{1 - 4\boldsymbol{\Omega}^2}R_n\right)} \qquad (35)$$

Next state the first order equation for ideal cloak parameters, equation (36):

$$\left(X_m^{(1)}\right)' + X_m^{(1)}\left(\frac{q(R_n + A)}{R_n} 2X_m^{(0)} + \frac{1}{R_n + A}\right) + \left(-2\frac{q(R_n + A)}{R_n}\left(X_m^{(0)}\right)^2\right)$$
$$+ \frac{1}{R_n + A}\left(2X_m^{(0)}\left(\frac{2(R_n + A)^2}{R_n^2} + 1\right)\right)$$
$$+ \frac{1}{q}\left(-\frac{A}{(R_n + A)R_n^2} - \frac{m^2 2}{(R_n + A)R_n}\right) = 0$$

Solution (34) can be substituted into equation (36) to find a solution for $X_m^{(1)}$, using the integrating factor method which leads to the following:

$$X_m^{(1)} = \frac{(1+A)}{(z+A')J_m^2(z)}(I_1 + I_2 + I_3 + I_4 + I_5) \quad (37)$$

Where:

$$I_1 = 2\int (J_m'(z))^2 z dz \quad (38)$$

$$I_2 = -2\int J_m'(z) J_m(z) \frac{z}{z+A'} dz \quad (39)$$

$$I_3 = -4\int J_m'(z) J_m(z) \frac{z+A'}{z} dz \quad (40)$$

$$I_4 = A'\int \frac{1}{z^2} J_m^2(z) dz \quad (41)$$

$$I_5 = 2m^2\int \frac{1}{z} J_m^2(z) dz \quad (42)$$

Using Bessel identities (Gradshteyn & Ryzhik, 2000), it is possible to solve these integrals and express the following solution for $X_m^{(1)}$, equation (43):

$$X_m^{(1)} = \frac{(1+A)}{z+A'}\left\{z^2\left(\frac{J_{m+1}^2(z)}{J_m^2(z)} - \frac{J_{m+2}(z)}{J_m(z)}\right) - \left(\frac{A'(1-4m)}{z(1-2m)} + 2 - 2m + \frac{2A'z}{4m^2-1} + \frac{z}{z+A'}\right)\right.$$
$$\left. + \frac{2A'}{4m^2-1}\frac{J_{m+1}(z)}{J_m(z)}\left(z\frac{J_{m-1}(z)}{J_m(z)} + 1\right) + \frac{A'}{J_m^2(z)}\int \frac{J_m^2(z)}{(z+A')^2} dz\right\}$$

Noting that the final integral $\int \frac{J_m^2(z)}{(z+A')^2} dz$ does not appear to have a solution and will be evaluated numerically.

Now that $X_m^{(0)}$ and $X_m^{(1)}$ have been found, expansion (29) can be solved at $R_n = 1$ and matched to boundary conditions (27) to yield solutions for the scattering coefficient $A_m$, enabling evaluation of the cloak performance.

### 2.1.4 Numerical solution

As a means of verifying the result produced by the perturbation expansion and matching solution, equation (21) will be solved numerically using an ODE solver in MATLAB. This method requires an input range of integration, with an initial condition defined at the beginning of that range. In the context of the rotating cloak in the frequency domain, this initial condition equates to a boundary condition at the inner cloak radius. Therefore, define a new, Neumann (hard wall) boundary condition at the cloak inner radius such that:

$$X_m(0) = 0 \tag{44}$$

Noting that this boundary condition can be added without making any difference to the stationary cloak performance, but is usually disregarded as unnecessary in the literature, see Cummer et al., (2008, p. 4) for more details.

To produce a valid solution, it will be necessary to solve equation (21) for the near cloak parameters, rather than the ideal cloak parameters. This is because the solver requires that the initial condition, in this case $X_m(0) = 0$, aligns with minimum value of the range of integration. Using the ideal cloak parameters at this value introduces a singularity, causing the method to fail. By introducing the small parameter $\delta$ in the near cloak formulation, this problem is avoided, without causing a significant difference to the cloaking behaviour. The perturbation expansion is not affected by this singularity issue, and can be evaluated for ideal parameters.

For a near cloak (Norris, 2008), the normalised, non-dimensional material parameters will be:

$$\bar{\rho}_r^{-1} = \frac{R_n + A_\delta}{R_n + A}, \qquad \bar{\rho}_\theta^{-1} = \frac{R_n + A}{R_n + A_\delta}, \qquad \lambda = \left(\frac{1 + A_\delta}{1 + A}\right)^2 \frac{(R_n + A)}{R_n + A_\delta} \tag{45}$$

Where $A_\delta = \frac{\delta}{(b-\delta)}$ and $R_n = \frac{R}{b-a}$

The MATLAB solver, ode23, follows a modified Rosenbrock formula order 2 scheme (MATLAB, 2024).

### 3.1 Results

### 3.1.1 Scattering amplitude

The results show a reasonable agreement (within 5%) between the perturbation expansion and numerical solution over the range of applicability, which suggests the perturbation expansion can accurately describe rotating cloak behaviour. In general, the results show that for slow speeds of rotation chosen where $\boldsymbol{\Omega}$ and $|\epsilon_m| \ll 1$, and for small cloaks (smaller than incident wavelength), that scattering caused by a rotating cloak is much less than the bare object alone.

The polar plots in Fig. 3. and Fig. 4. plot the absolute scattering amplitude $|f(\theta)|$, where:

$$f(\theta) = \sum_{m=-\infty}^{+\infty} A_m e^{im\theta} \tag{46}$$

also known as scattering form factor, which is an indication of the far field behaviour (Censor & Aboudi, 1971; Cai & Sánchez-Dehesa, 2007). For all plots, an incident plane wave, unit amplitude is incoming from the left. Legend: Ideal cloak perturbation expansion (magenta); Near cloak numerical solution (blue); uncloaked rigid core (black). All results truncate the range of harmonics to $-17 \leq m \leq 17$ .

Fig. 3. (a)-(h) Show that rotation causes scattering, with scattering increasing as rotational speed increases. But for this small sized cloak ($kb = 1.35$), the scattering is less than the scattering caused by the uncloaked rigid core ($ka = 0.21$). Fig. 3. (a)-(b) Show that very slow speeds of rotation cause near zero scattering. Overall, a small rotating cloak still offers a reduced cloaking performance whilst rotating at slow speed.

Fig. 4. (a)-(h) Tests a larger diameter cloak to Fig. 3., whilst the other variables like rotational speed, incident wavelength and radius of cloaked region remain the same as Fig. 3. It is evident that a larger diameter rotating cloak produces more scattering effects. Fig. 4. (g)-(h) show that with increased rotational speed, the magnitude of the scattering caused by rotation is comparable to the uncloaked rigid core.

The results from Fig. 3. and Fig. 4. suggest that a good cloaking performance can be preserved in cases where a rotating cloak is small compared to the incident wavelength and rotating slowly.

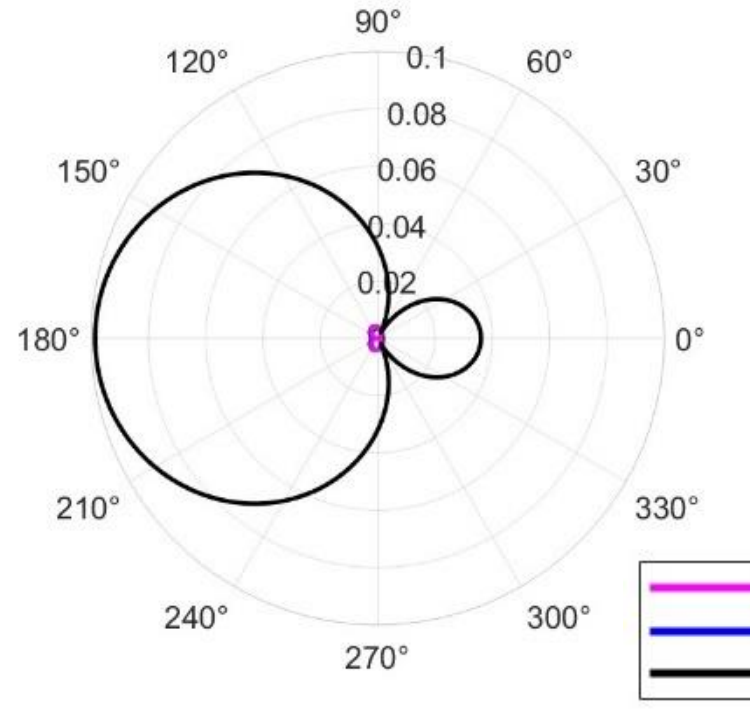

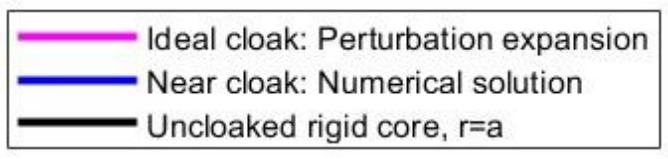

Fig. 3. (a) Scattering amplitudes for uncloaked rigid core and rotating cloaks: $ka = 0.21, kb = 1.45, \boldsymbol{\Omega} = 0.0032$

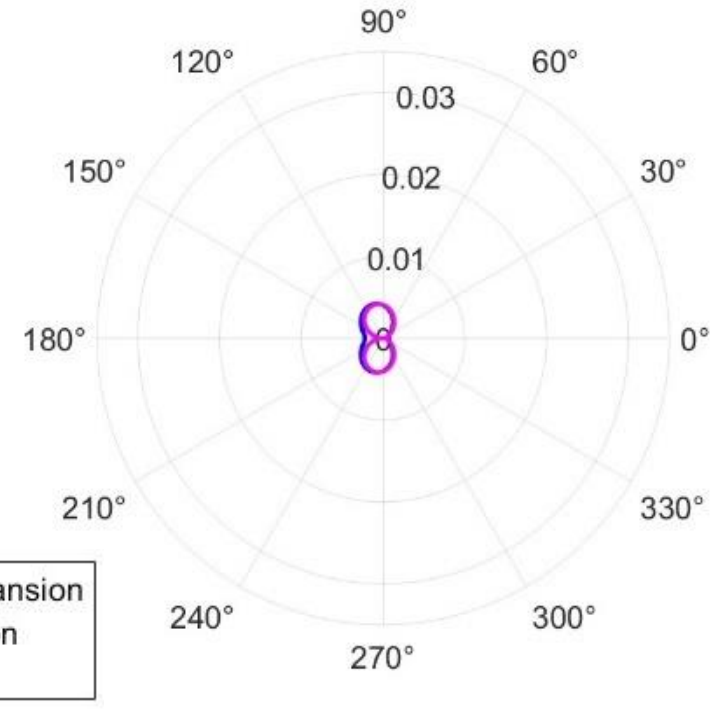

Fig. 3. (b) Scattering amplitudes for rotating cloaks: $ka = 0.21, kb = 1.45, \boldsymbol{\Omega} = 0.0032$

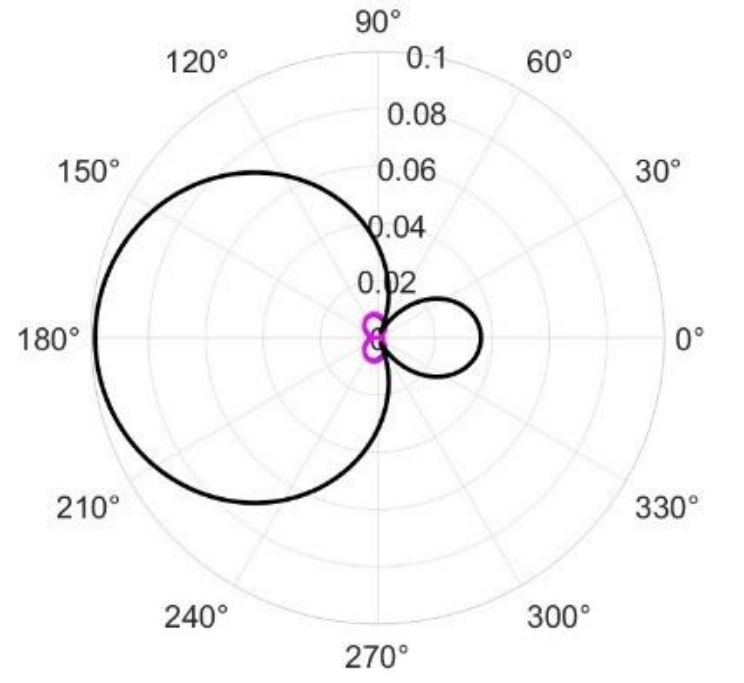

Fig. 3. (c) Scattering amplitudes for uncloaked rigid core and rotating cloaks: $ka = 0.21, kb = 1.45, \boldsymbol{\Omega} = 0.0065$

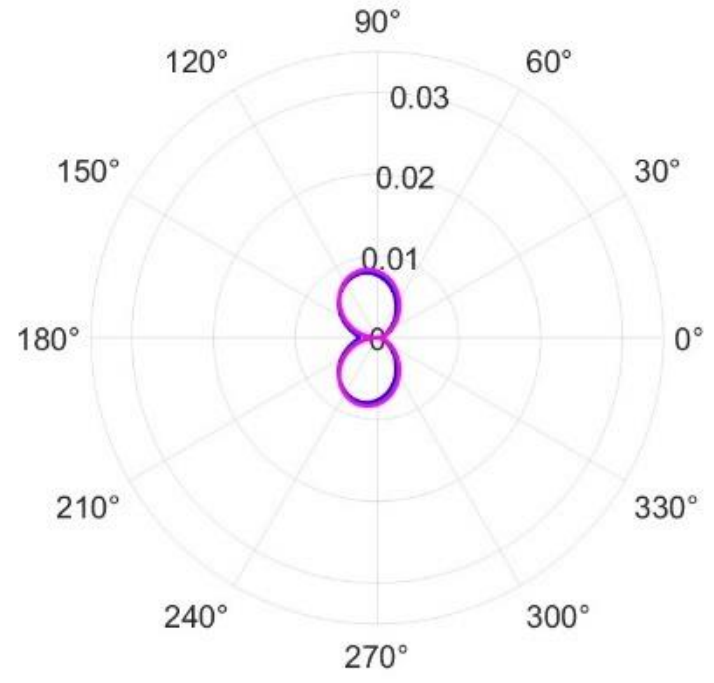

Fig. 3. (d) Scattering amplitudes for rotating cloaks: $ka = 0.21, kb = 1.45, \boldsymbol{\Omega} = 0.0065$

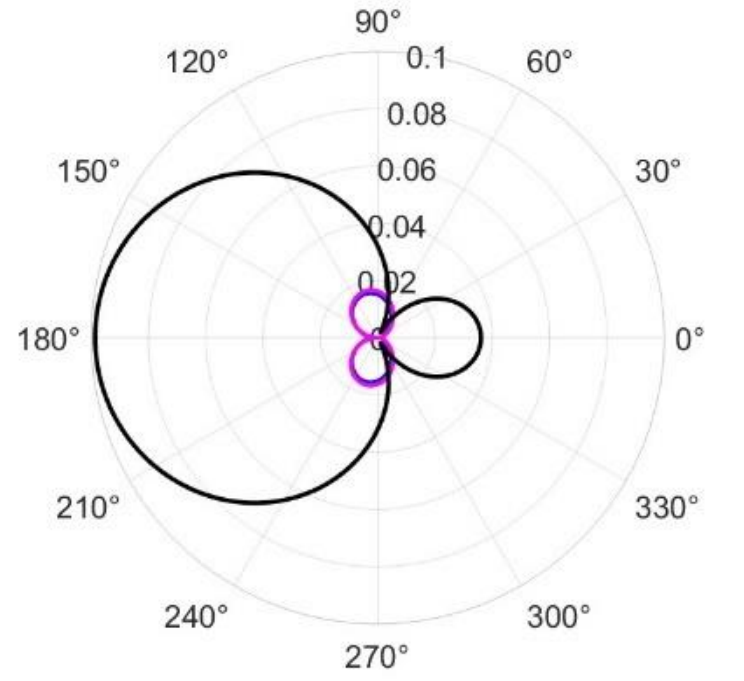

Fig. 3. (e) Scattering amplitudes for uncloaked rigid core and rotating cloaks: $ka = 0.21, kb = 1.45, \boldsymbol{\Omega} = 0.013$

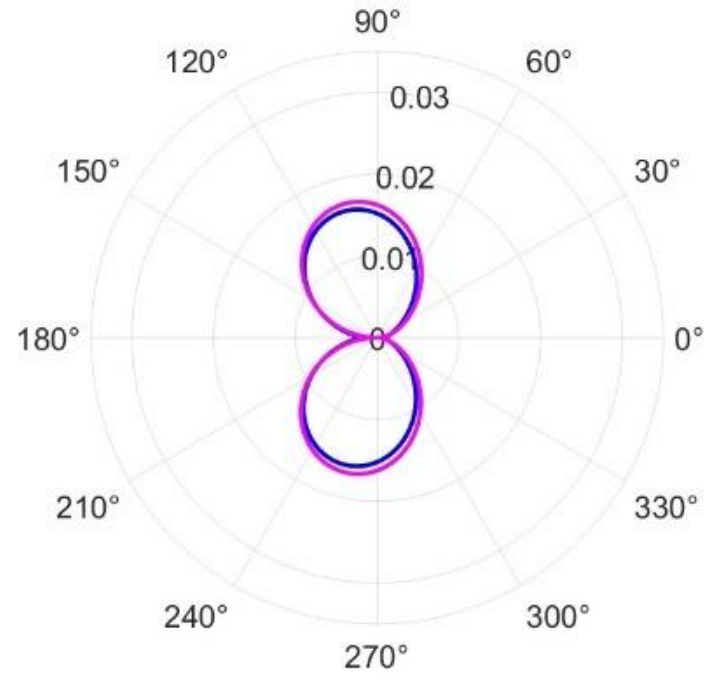

Fig. 3. (f) Scattering amplitudes for rotating cloaks: $ka = 0.21, kb = 1.45, \boldsymbol{\Omega} = 0.013$

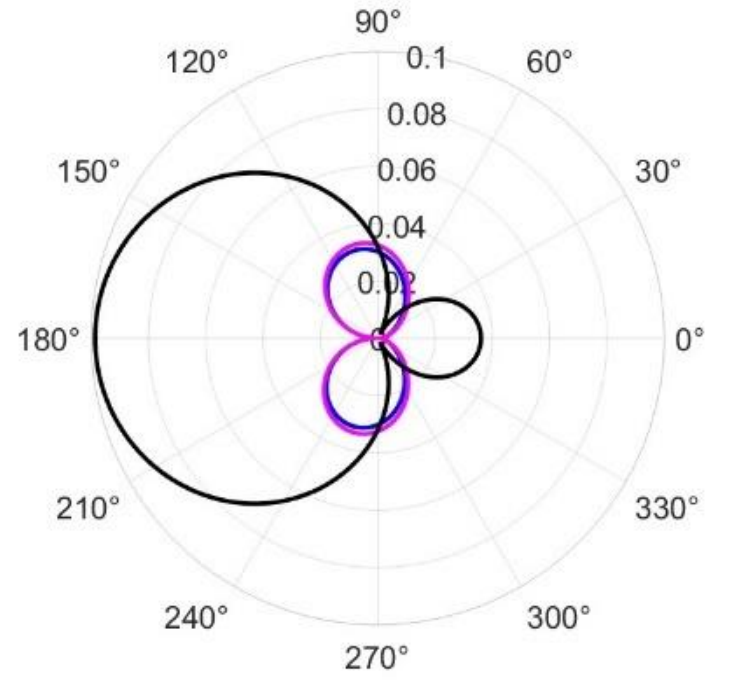

Fig. 3. (g) Scattering amplitudes for uncloaked rigid core and rotating cloaks $ka = 0.21, kb = 1.45, \boldsymbol{\Omega} = 0.026$

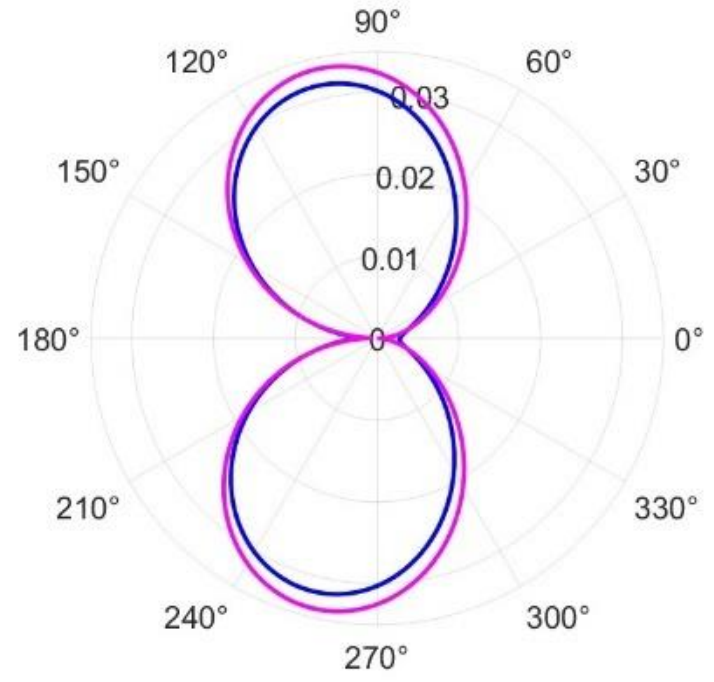

Fig. 3. (h) Scattering amplitudes for rotating cloaks: $ka = 0.21, kb = 1.45, \boldsymbol{\Omega} = 0.026$

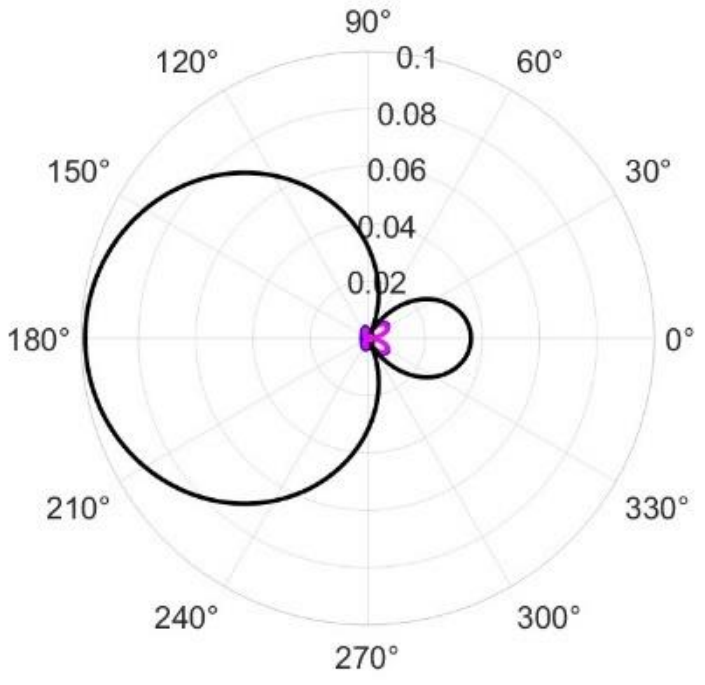

Fig. 4. (a) Scattering amplitudes for uncloaked rigid core and rotating cloaks $ka = 0.21, kb = 2.35, \boldsymbol{\Omega} = 0.0032$

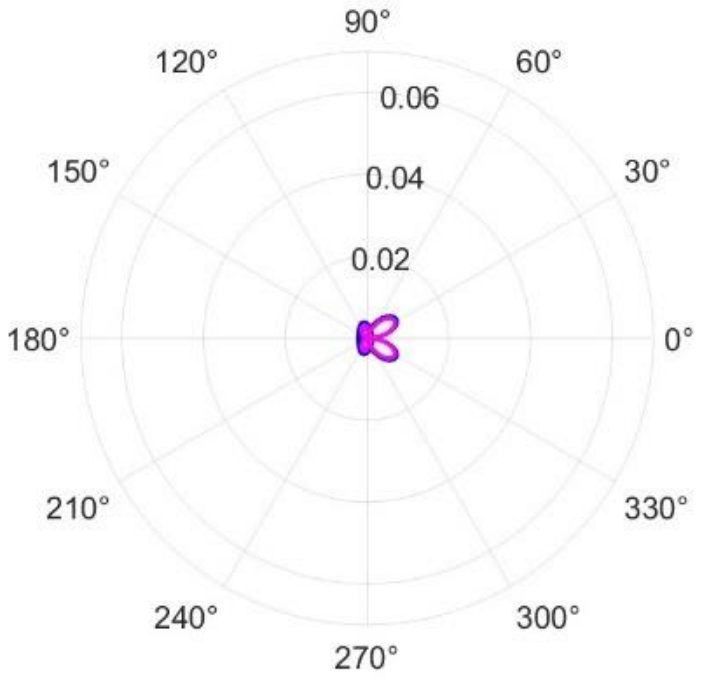

Fig. 4. (b) Scattering amplitudes for rotating cloaks: $ka = 0.21, kb = 2.35, \boldsymbol{\Omega} = 0.0032$

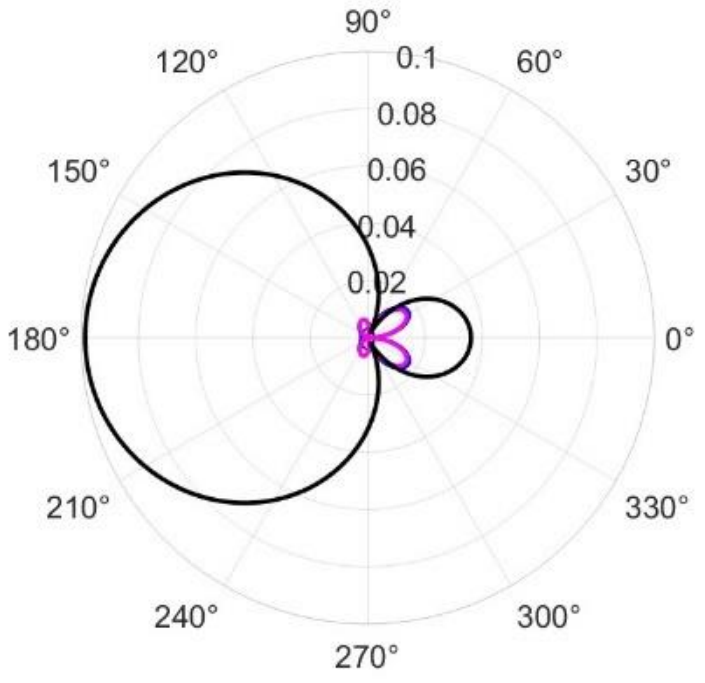

Fig. 4. (c) Scattering amplitudes for uncloaked rigid core and rotating cloaks: $ka = 0.21, kb = 2.35, \boldsymbol{\Omega} = 0.0065$

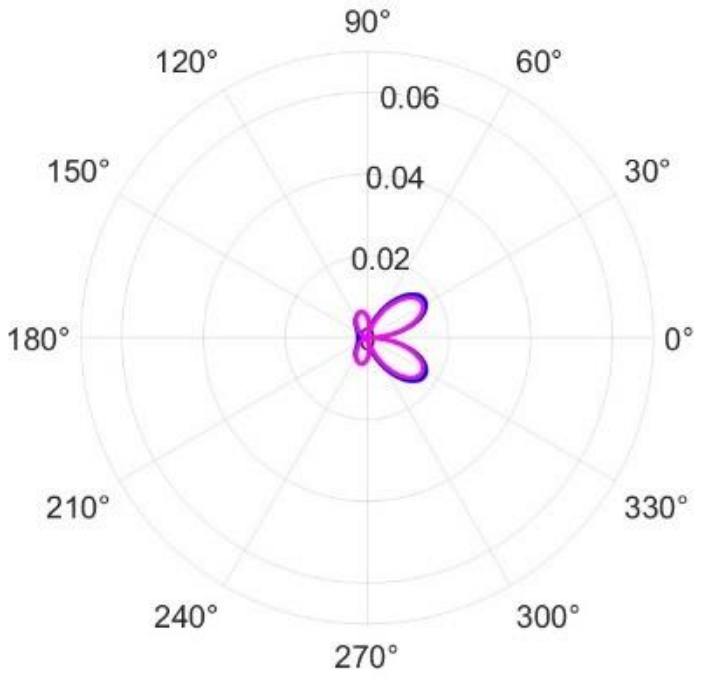

Fig. 4. (d) Scattering amplitudes for rotating cloaks: $ka = 0.21, kb = 2.35, \boldsymbol{\Omega} = 0.0065$

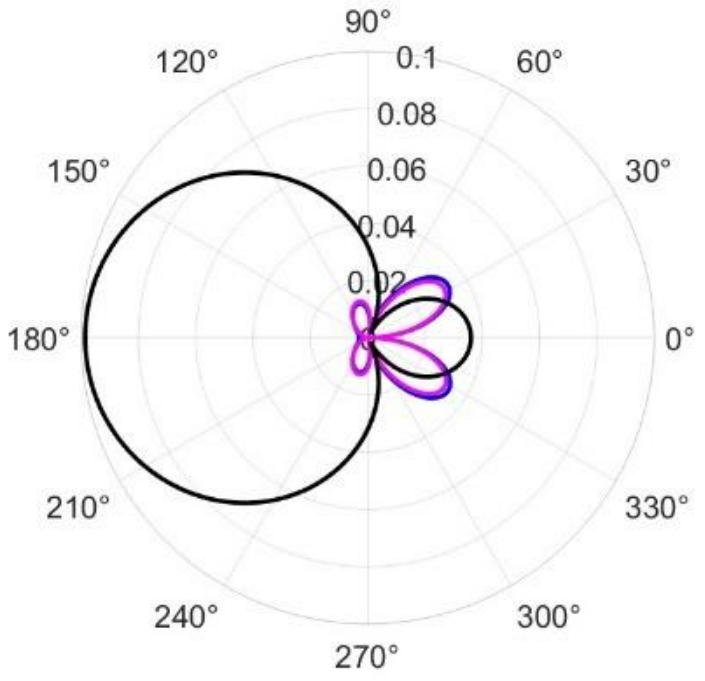

Fig. 4. (e) Scattering amplitudes for uncloaked rigid core and rotating cloaks: $ka = 0.21, kb = 2.35, \boldsymbol{\Omega} = 0.013$

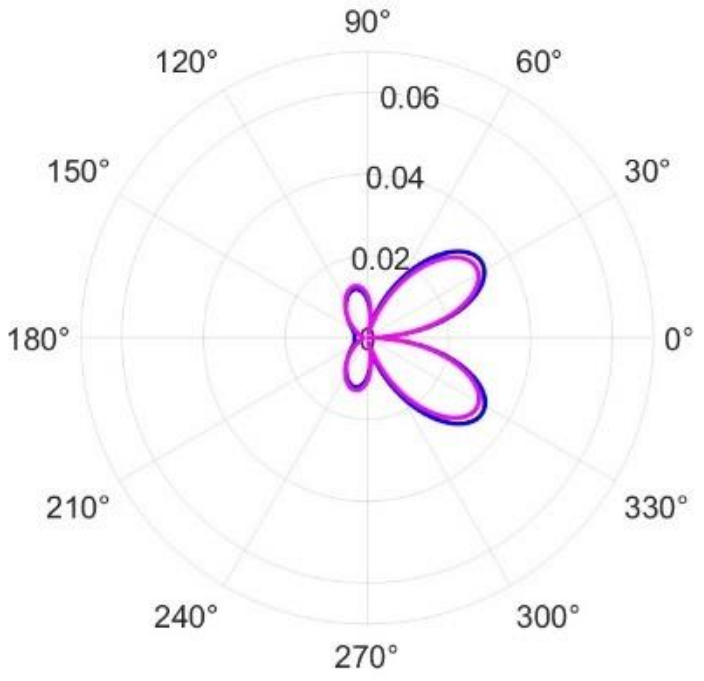

Fig. 4. (f) Scattering amplitudes for rotating cloaks: $ka = 0.21, kb = 2.35, \boldsymbol{\Omega} = 0.013$

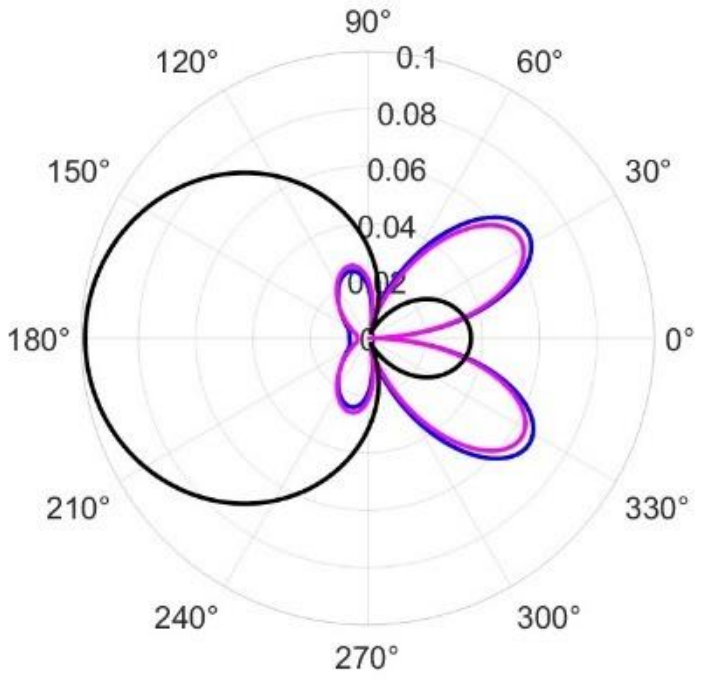

Fig. 4. (g) Scattering amplitudes for uncloaked rigid core and rotating cloaks: $ka = 0.21, kb = 2.35, \boldsymbol{\Omega} = 0.026$

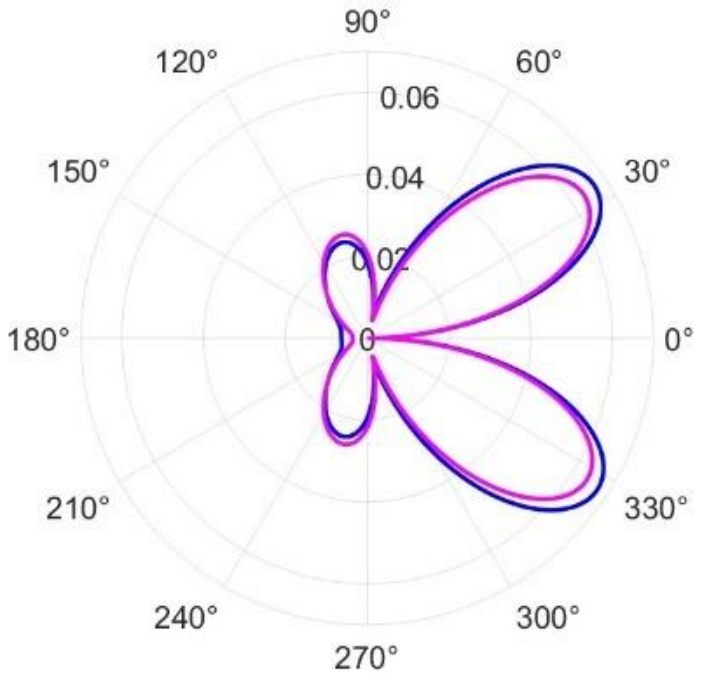

Fig. 4. (h) Scattering amplitudes for rotating cloaks: $ka = 0.21, kb = 2.35, \boldsymbol{\Omega} = 0.026$

### 3.1.2 Scattering Cross Section

Fig. 5. (a)-(d) plot the ratio of Scattering Cross Sections:

$$\frac{\sigma_{cloak}}{\sigma_{rigid}} \tag{47}$$

Where Scattering Cross Section (SCS):

$$\sigma = \frac{4}{k}\sum_{m=-\infty}^{+\infty}|A_m|^2 \tag{48}$$

The SCS is the total energy transmitted by the wave in the far field and describes how strong the scattering object is (Cai & Sánchez-Dehesa, 2007; Farhat et al., 2020). By plotting the SCS ratio of the cloak to a rigid cylinder of the same diameter ($r = b$), it is possible to see how a rotating cloak performs against a rigid cylinder of the same size. If the ratio is one, then the rotating cloak is producing as much scattering as the rigid cylinder.

Fig. 5. (a)-(b) are plotted for two values of constant rotational speed, $\Omega$. The cloak dimensions remain unchanged, and the incident frequency increases. The plots show that generally the rotating cloak produces much less scattering that the rigid cylinder, however there is an increase in scattering for lower incident frequencies. As the incident wavelength becomes smaller, it starts to become comparable in size to the rotational speed and $\boldsymbol{\Omega}$ becomes larger, causing more scattering. This indicates that there may be some resonant responses for cloaks subject to low frequency incident waves, even at slow speeds of rotation. An increase in speed causes an increase in scattering for all values of incident frequency.

Fig. 5. (c)-(d) are plotted for two values of constant $\boldsymbol{\Omega} = \Omega/\omega$. The cloak dimensions remain unchanged, and the incident frequency increases. The plots show that for the low rotational speeds chosen, the scattered energy is extremely small compared to the object alone – at most around 2.5%. They show that an increase in the rotational speed increases the scattering for values of incident frequency.

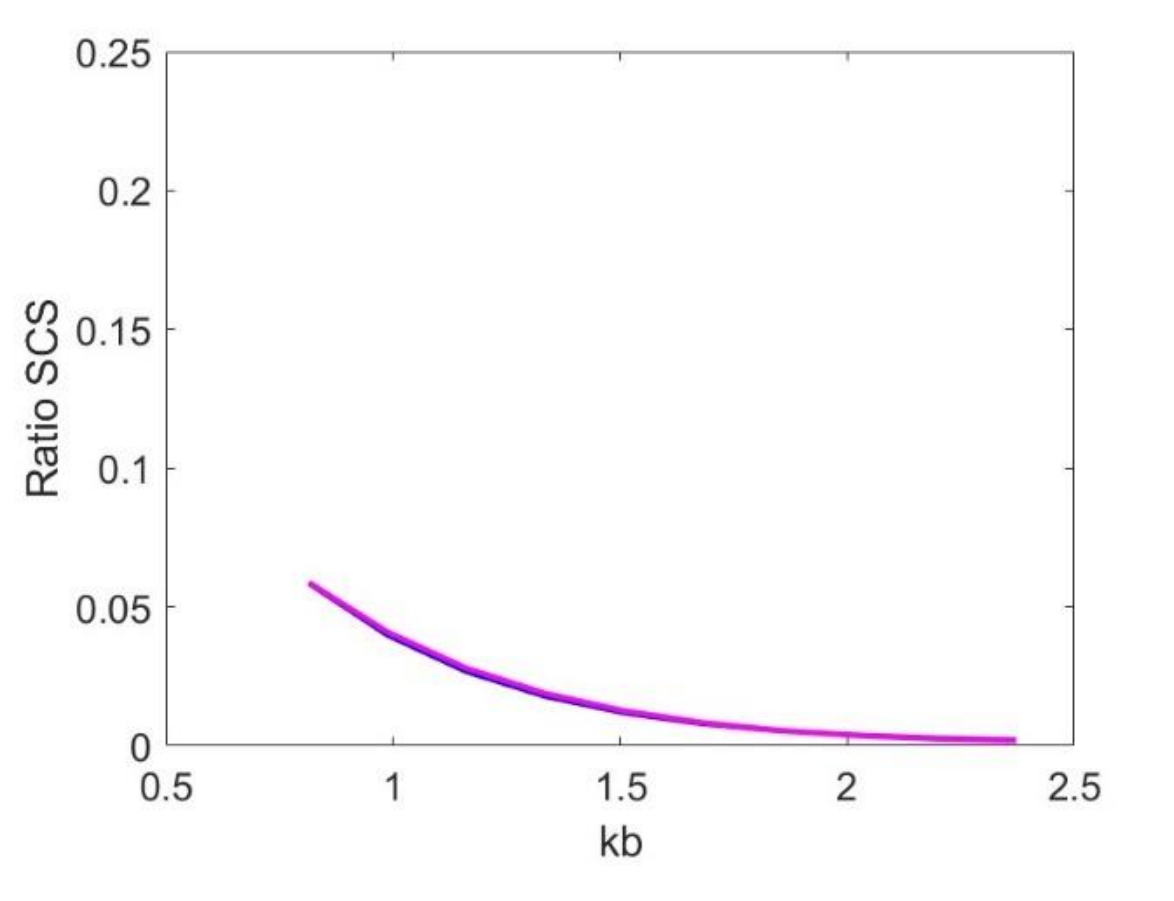

Fig. 5. (a) Ratio SCS rotating cloaks, rigid cylinder:

$a = 0.05\ m, b = 0.35\ m, \Omega = 9.13\ rad/s$

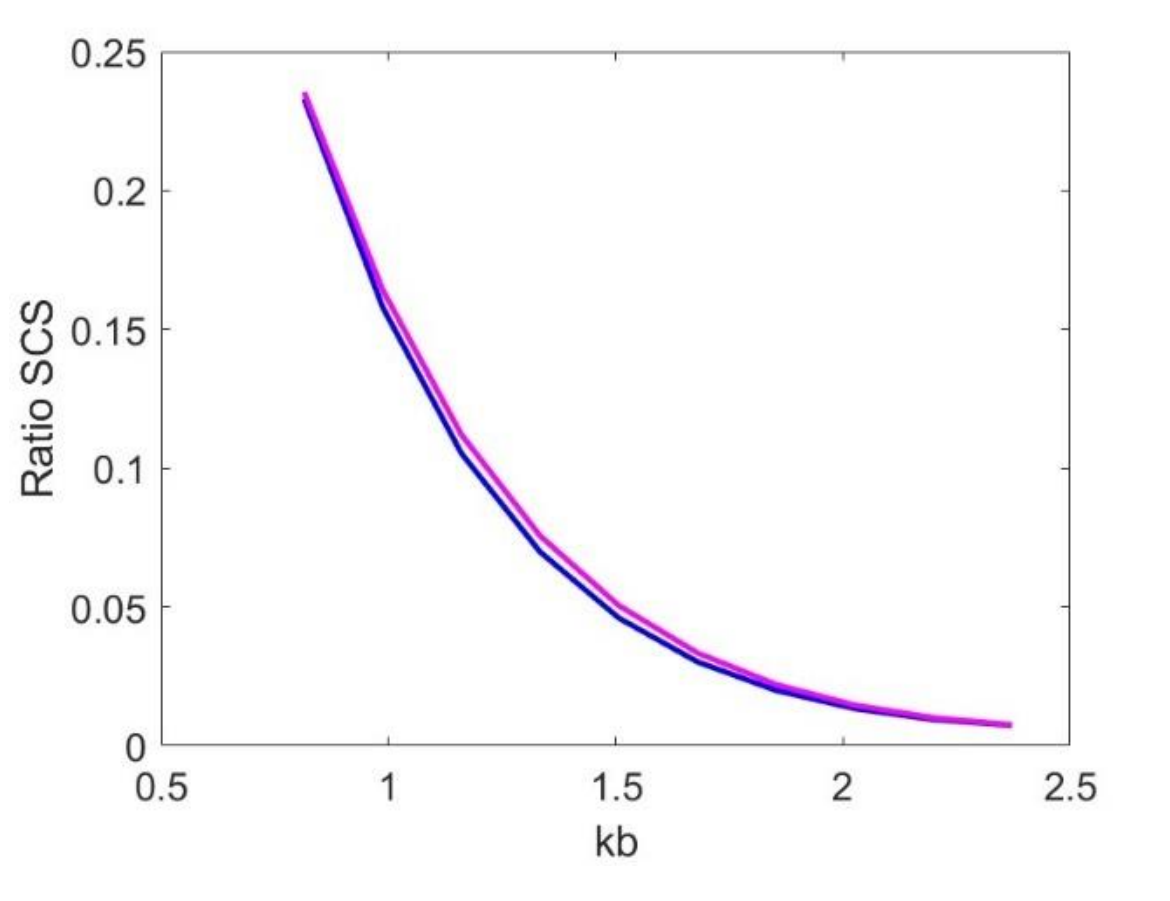

Fig. 5. (b) Ratio SCS rotating cloaks, rigid cylinder:

$a = 0.05\ m, b = 0.35\ m, \Omega = 18.3\ rad/s$

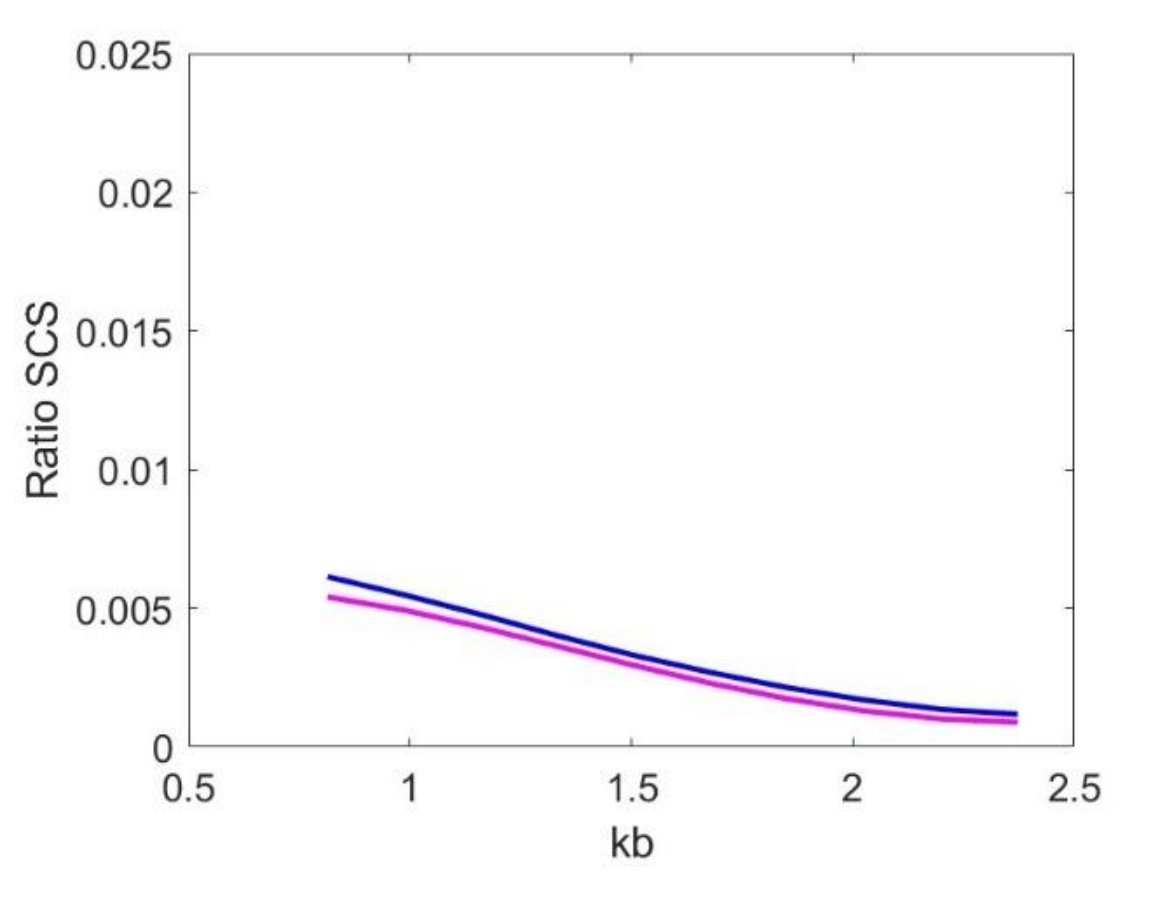

Fig. 5. (c) Ratio SCS rotating cloaks, rigid cylinder:

$a = 0.05\ m, b = 0.35\ m, \boldsymbol{\Omega} = 0.0035$

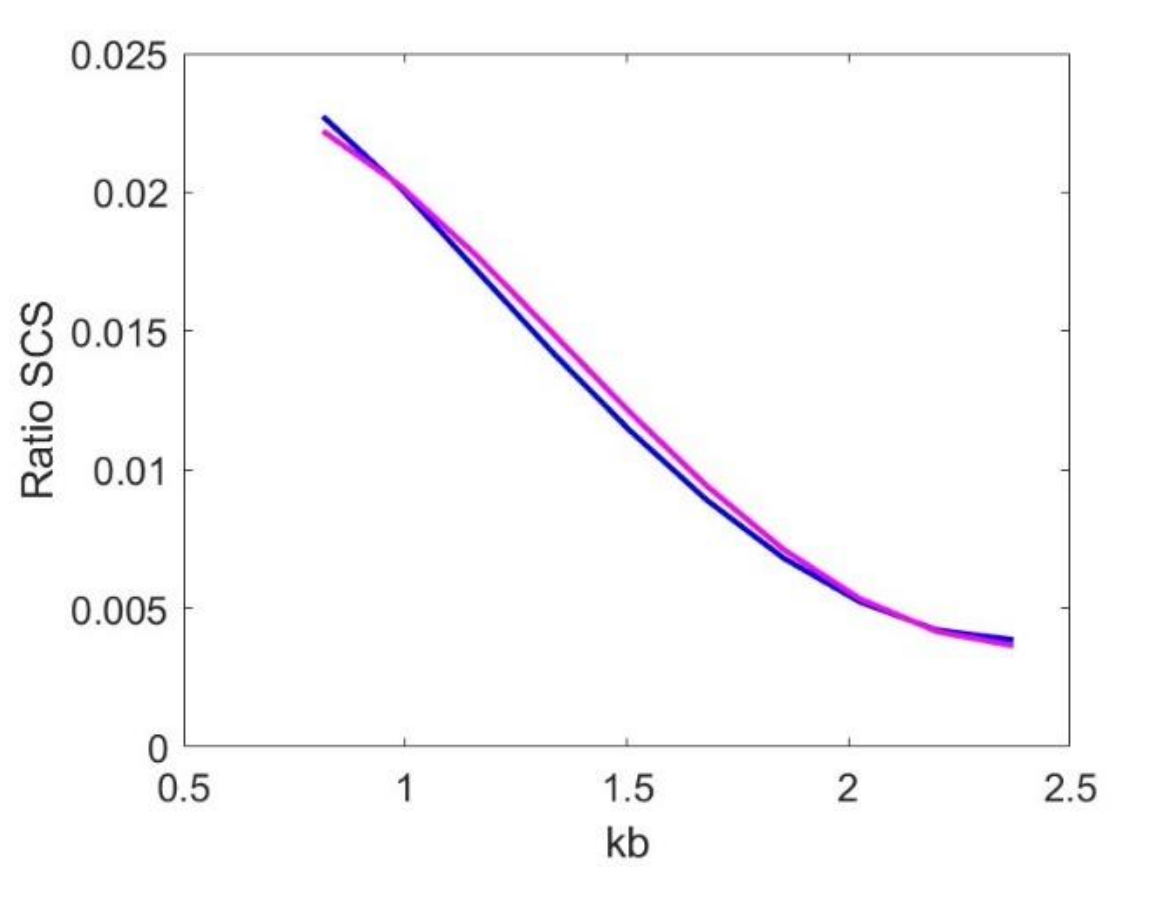

Fig. 5. (d) Ratio SCS rotating cloaks, rigid cylinder:

$a = 0.05\ m, b = 0.35\ m, \boldsymbol{\Omega} = 0.007$

### 3.1.3 Increasing cloaking region

Fig. 6. (a)-(h) plot the scattering amplitude for the rotating cloak: ideal cloak perturbation expansion, and near cloak numerical solution, for increasing values of cloaked region radius, $a$. In each plot, the size of the outer radius, $b$, remains unchanged, and the rotational speed remains unchanged. The plots show that increasing the cloaked area has a significant effect on the scattering, changing directionality and / or magnitude. The plots also show that as $a \to b/2$, the numerical solution and perturbation expansion start to show a discrepancy between results, no longer within 5%.

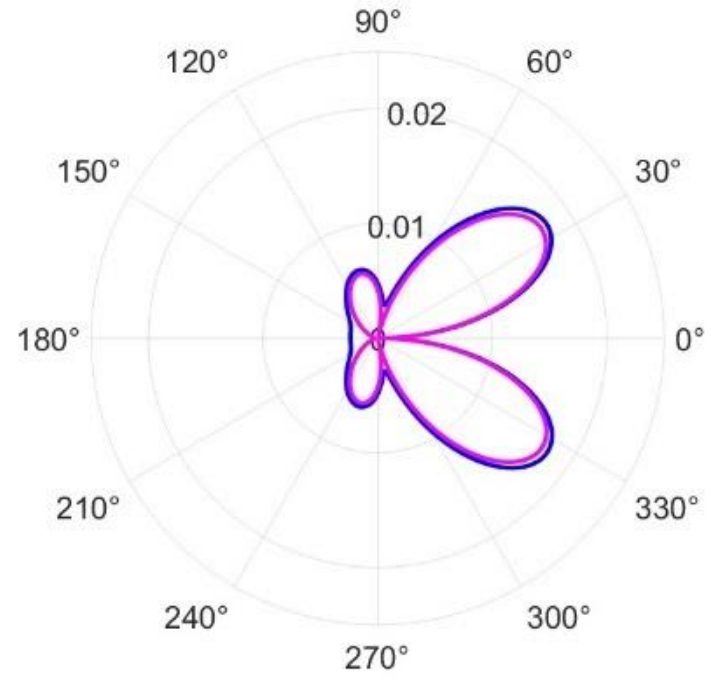

Fig. 6. (a) Scattering amplitudes for rotating cloaks:

$ka = 0.052, kb = 2.35, \boldsymbol{\Omega} = 0.0065$

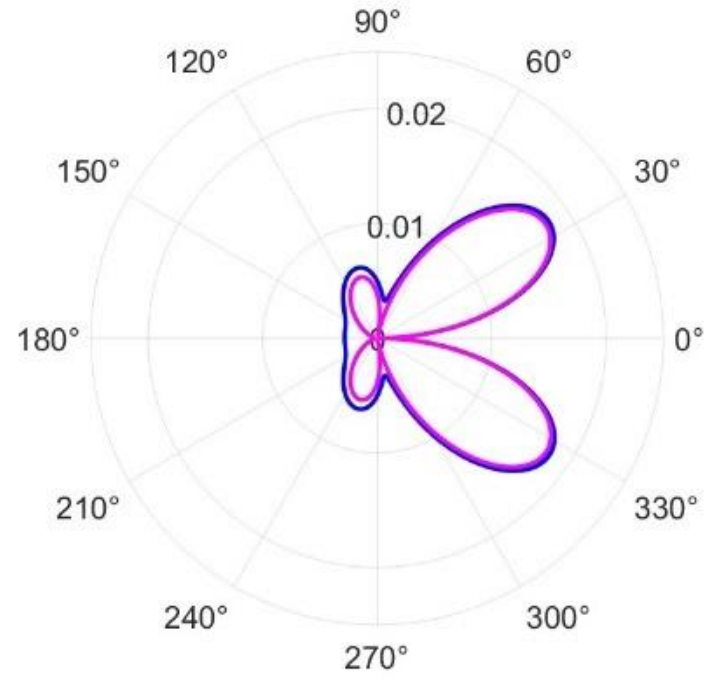

Fig.6. (b) Scattering amplitudes for rotating cloaks:

$ka = 0.1, kb = 2.35, \boldsymbol{\Omega} = 0.0065$

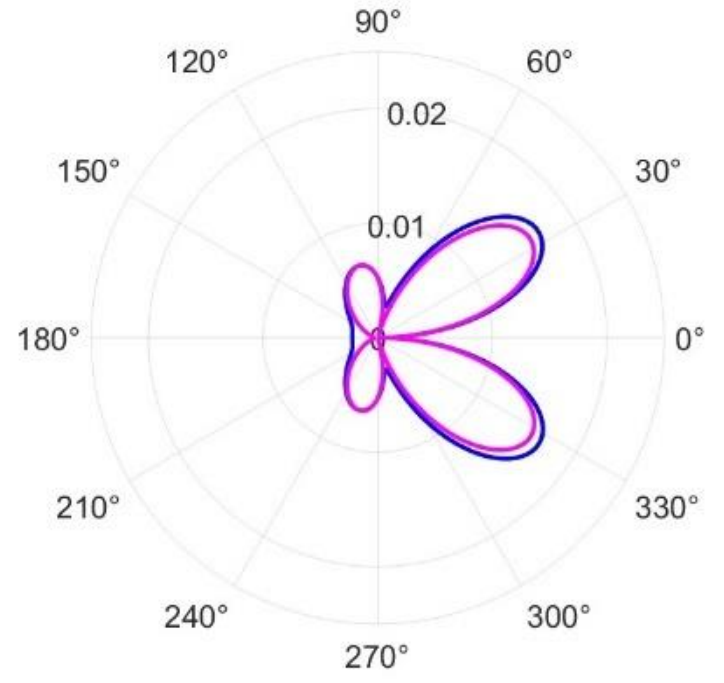

Fig. 6. (c) Scattering amplitudes for rotating cloaks:

$ka = 0.21, kb = 2.35, \boldsymbol{\Omega} = 0.0065$

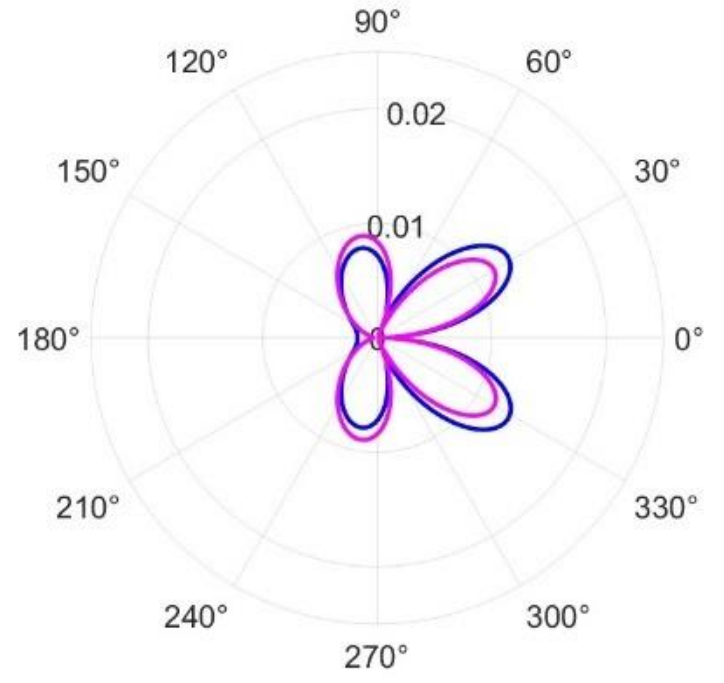

Fig. 6. (d) Scattering amplitudes for rotating cloaks:

$ka = 0.45, kb = 2.35, \boldsymbol{\Omega} = 0.0065$

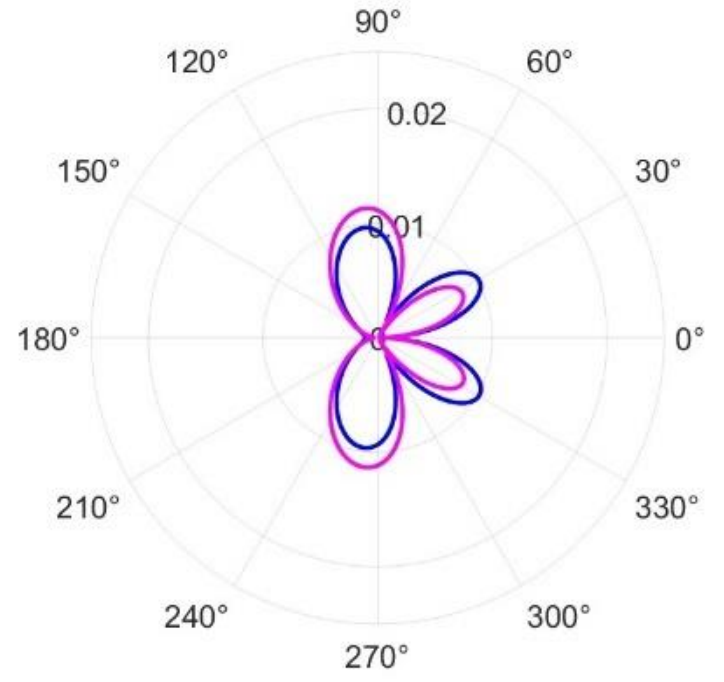

Fig. 6. (e) Scattering amplitudes for rotating cloaks:

$ka = 0.62, kb = 2.35, \boldsymbol{\Omega} = 0.0065$

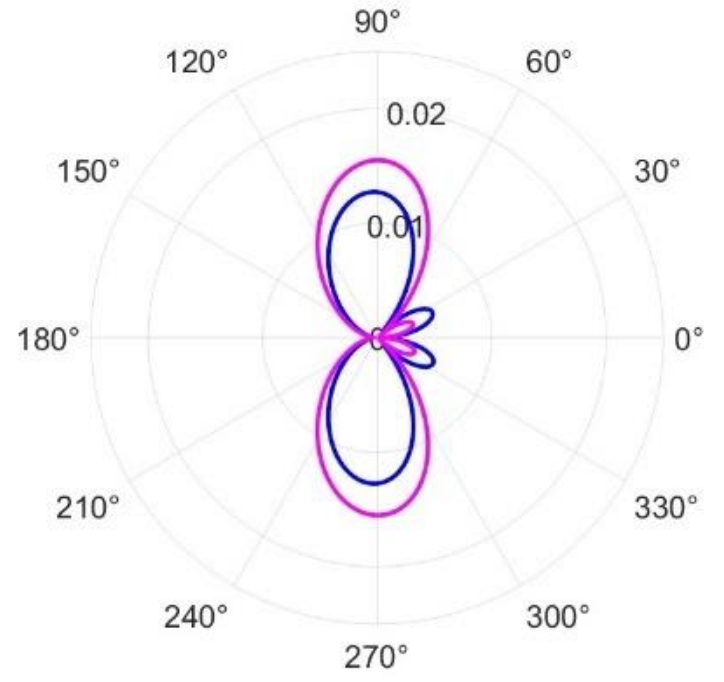

Fig. 6. (f) Scattering amplitudes for rotating cloaks:

$ka = 0.82, kb = 2.35, \boldsymbol{\Omega} = 0.0065$

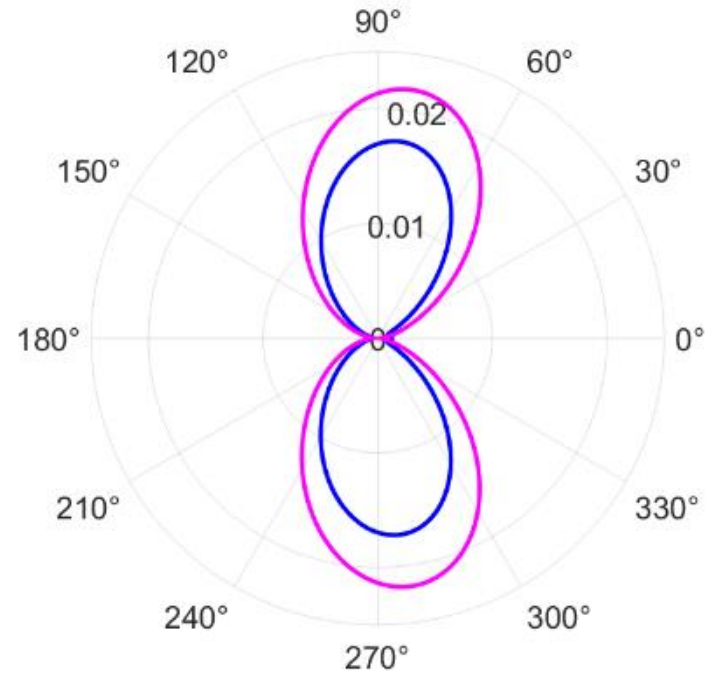

Fig. 6. (g) Scattering amplitudes for rotating cloaks:

$ka = 1.03, kb = 2.35, \boldsymbol{\Omega} = 0.0065$

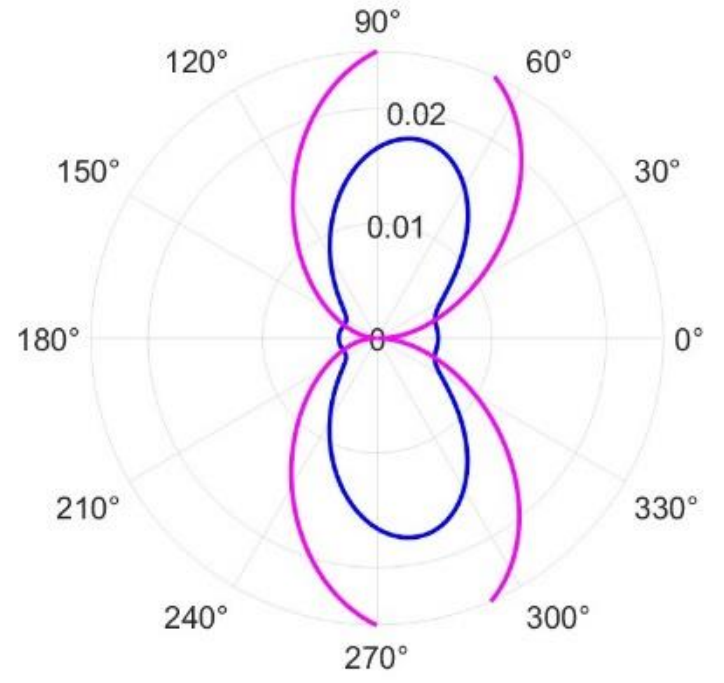

Fig. 6. (h) Scattering amplitudes for rotating cloaks:

$ka = 1.13, kb = 2.35, \boldsymbol{\Omega} = 0.0065$

### 3.1.4 Total pressure field

Fig. 7. (a)-(b) are two dimensional plots of the total pressure field: $J_m(kr) + A_m H_m(kr)$, plotted for $b \leq r \leq 5 \times wavelength; 0 < \theta \leq 2\pi$. Fig. 7. (a) shows that the scattering caused by an uncloaked rigid core interferes with the incident plane wave, making it easy to detect. Fig. 7. (b) shows that the rotating cloaked rigid core still offers a good cloaking performance despite rotating at a speed towards the higher end of the acceptable range of $\boldsymbol{\Omega}$. Plotting codes written by Oyvind Breivik, University of Bergen.

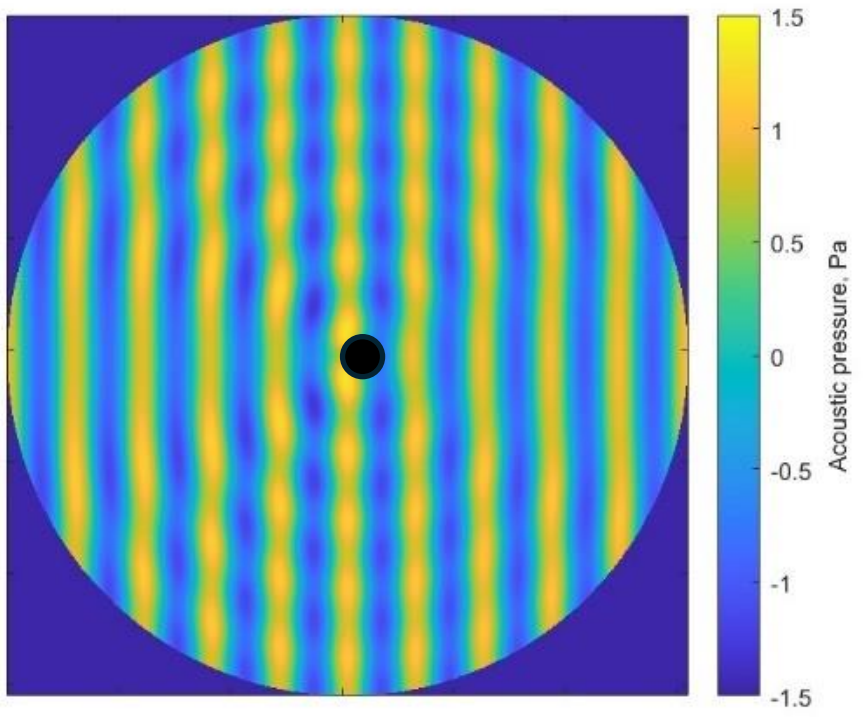

Fig. 7. (a) Total pressure field for uncloaked rigid core:

$ka = 1.13$

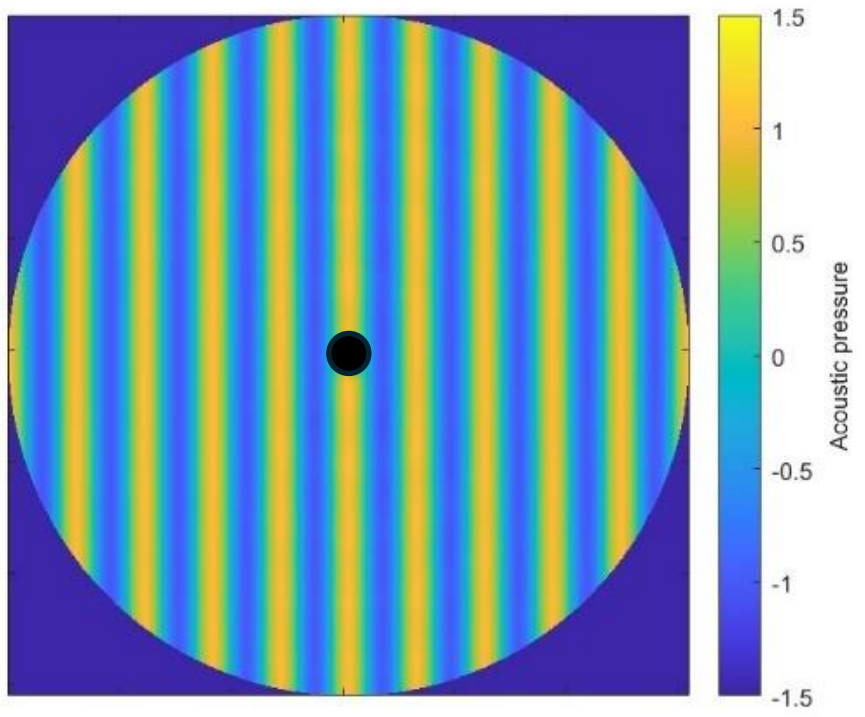

Fig. 7. (b) Total pressure field for rotating cloaked core:

$kb = 2.35, ka = 1.13, \boldsymbol{\Omega} = 0.026$

### 3.1.5 Outgoing pressure wave

Finally, Fig. 8. (a)-(b) show the pressure wave ($A_m H_m(kr)$) outgoing from the cloak surface, $r = b$. These plots highlight the dipole behaviour of the rotating cloak, increasing in strength as rotational speed increases.

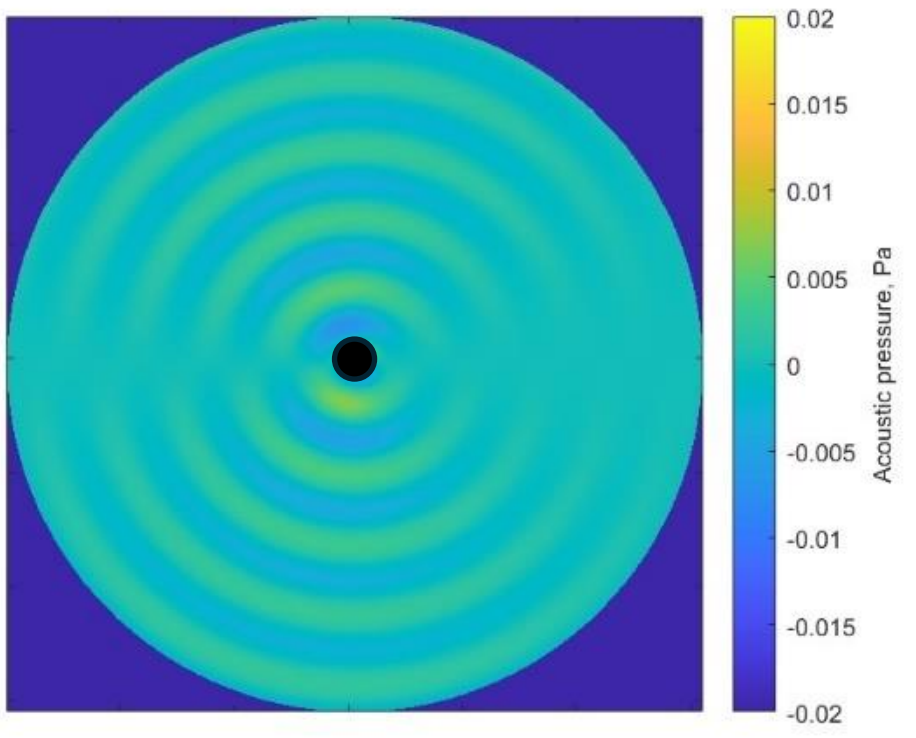

Fig. 8. (a) Outgoing pressure wave:

$ka = 0.21, kb = 1.45, \boldsymbol{\Omega} = 0.013$

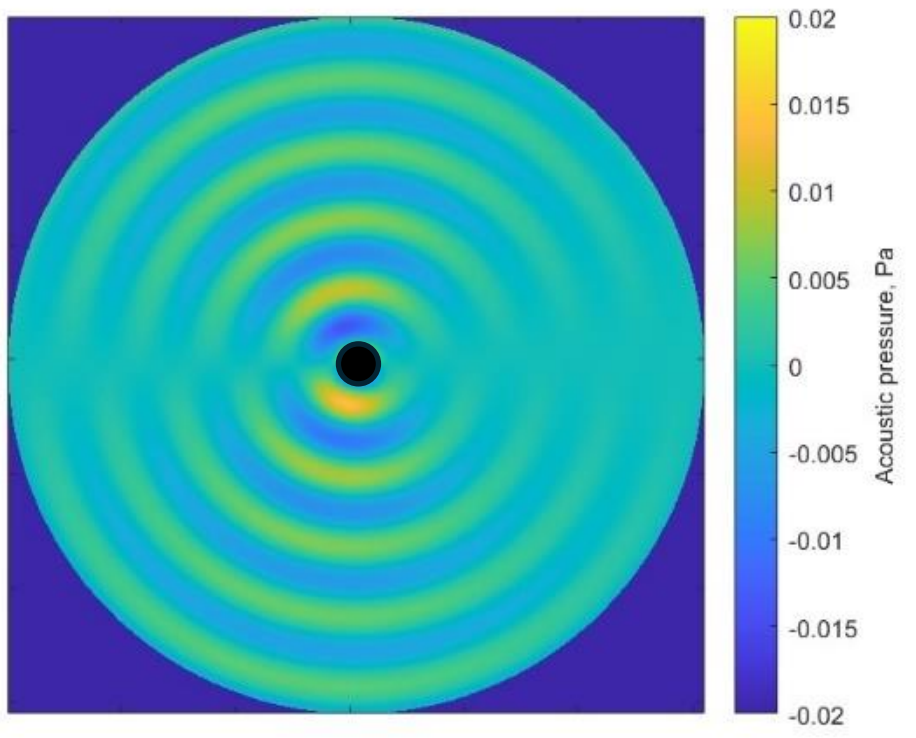

Fig. 8. (b) Outgoing pressure wave:

$ka = 0.21, kb = 1.45, \boldsymbol{\Omega} = 0.026$

### 4.1 Discussion

The results have demonstrated that, to varying degrees, rotation causes a reduction in cloaking performance. This matches expectations, as the form of the differential equation governing wave propagation no longer holds form invariance as required for transformation acoustics. This result shows agreement with findings in the literature studying the cloak in a flow, such as Zhong and Huang (2013), Iemma (2016) and Colombo (2024), where cloaking performance is reduced by a uniform mean flow, with increased scattering for higher flow speeds. It also finds agreement with the study of a rotating cloak in the electromagnetic case, for which it was shown that cloaking performance is reduced by rotation and that high speeds of rotation cause large scattering effects (Hasanpour Tadi & Shokri, 2024).

Despite the reduction in performance, a promising result is that the overall scattering produced by the rotating cloak is relatively small compared to the uncloaked rigid core when the incident wavelength is large compared to the cloak and the rotational speed is low. This suggests that there may be a range of aeronautical applications for which the cloak can be successfully used, such as small airframe elements with limited rotation. Outside of this range, it is likely that modified cloaking parameters need to be discovered to successfully deal with high rotational speeds and small wavelengths. Of course, this study has not considered the full range of aerodynamic effects such as viscosity and turbulence that would be present in a realistic scenario, and these would need to be further investigated.

In question is whether parameters that would achieve perfect cloaking whilst rotating actually exist. These currently seem difficult to achieve, given that the governing wave equation (17), (21) introduces modal speeds of wave propagation. Parameters would need to simultaneously cancel these modes or cancel only the dominant modes. Farhat et al., (2020), consider a related problem, aiming to achieve cloaking in a homogeneous fluid. By considering the two dominant modes contributing to scattering, a cloaking effect is realised by two contrarotating cylinders. This solution would probably be impractical in a realistic scenario. Future work will investigate the existence of improved cloaking parameters whilst rotating.

The study has successfully developed a perturbation method which calculates the scattering caused by a rotating, anisotropic, graded, cylindrical fluid region within a limited range of conditions. The perturbation method offers some advantages over a numerical solution in that it is stable, not suffering from issues with singularities, and computationally inexpensive, able to produce a result much faster than the numerical solution. These qualities naturally make it a useful tool in seeking improved cloaking parameters, especially when combined with an optimisation method. As noted in the results, discrepancies appear between the perturbation solution and numerical solution as the size of the cloaked region increases. It is not clear why this is the case, and further validation by means of CFD or FEM would assist in establishing the more accurate solution and range of validity.

The analytical tools developed in this study have implications beyond the immediate problem. The generalised wave equation can be used to discover insights into the effect of rotation on other anisotropic metamaterials, which may be a useful avenue towards modelling complicated propagation features inside aero-engine ducts (Palma & Iemma, 2023). This will be a focus of future research.

### 5.1 Conclusion

In summary, this study has derived the general wave equation for a rotating, anisotropic, graded, cylindrical fluid region. This equation is used to model a rotating acoustic invisibility cloak, with parameters defined by the transformation acoustics method. Solutions to the equation have been found through a semi-analytical perturbation method and numerical solution. The solutions show a reasonable agreement (within 5%) and are valid for cases of slow rotational speed, and for small cloaks, with small cloaking regions. The results show that under these conditions, a good cloaking performance still exists compared to the uncloaked rigid core. Future work will focus on validation of these results using CFD or FEM and extending the methods here to consider higher speeds of rotation.

### Acknowledgments

This research was funded by the University of Salford Widening Participation scholarship.

**References**

Advisory Council for Aviation Research and Innovation in Europe. (2024). *Goals.* https://www.acare4europe.org/acare-goals/

Cai, L. W., & Sánchez-Dehesa, J. (2007). Analysis of Cummer–Schurig acoustic cloaking. *New journal of physics*, *9*(12), 450. doi: 10.1088/1367-2630/9/12/450

Chen, H., & Chan, C. T. (2007). Acoustic cloaking in three dimensions using acoustic metamaterials. *Applied physics letters*, *91*(18), 183518. doi: 10.1063/1.2803315

Chen, H., & Chan, C. T. (2010). Acoustic cloaking and transformation acoustics. *Journal of Physics D: Applied Physics*, *43*(11), 113001. doi: 10.1088/0022-3727/43/11/113001

Colombo, G., Palma, G., & Iemma, U. (2024). Numerical Aeroacoustic Assessment of a Metacontinuum Device Impinged by a Laser-Generated Sound Source. *Journal of Engineering*, *2024*(1), 5421663. doi: 10.1155/2024/5421663

Cummer, S. A., Popa, B. I., Schurig, D., Smith, D. R., Pendry, J., Rahm, M., & Starr, A. (2008). Scattering theory derivation of a 3D acoustic cloaking shell. *Physical review letters*, *100*(2), 024301. doi: 10.1103/PhysRevLett.100.024301

Cummer, S. A. (2013). Transformation Acoustics. In Craster, R. V., & Guenneau, S. (Eds.), *Acoustic metamaterials*. (pp. 197-217). doi: 10.1007/978-94-007-4813-2_8

Farhat, M., Guenneau, S., Alù, A., & Wu, Y. (2020). Scattering cancellation technique for acoustic spinning objects. *Physical Review B, 101*(17), 174111. doi: 10.1103/PhysRevB.101.174111

García-Meca, C., Carloni, S., Barceló, C., Jannes, G., Sánchez-Dehesa, J., & Martínez, A. (2013). Analogue transformations in physics and their application to acoustics. *Scientific reports*, *3*(1), 2009. doi: 10.1038/srep02009

García-Meca, C., Carloni, S., Barceló, C., Jannes, G., Sánchez-Dehesa, J., & Martínez, A. (2014a). Space–time transformation acoustics. *Wave Motion*, *51*(5), 785-797. doi: 10.1016/j.wavemoti.2014.01.008

García-Meca, C., Carloni, S., Barceló, C., Jannes, G., Sánchez-Dehesa, J., & Martínez, A. (2014b). Analogue transformation acoustics and the compression of spacetime. *Photonics and nanostructures-fundamentals and applications*, *12*(4), 312-318. doi: 10.1016/j.photonics.2014.05.001

Gokhale, N. H., Cipolla, J. L., & Norris, A. N. (2012). Special transformations for pentamode acoustic cloaking. *The Journal of the Acoustical Society of America*, *132*(4), 2932-2941. doi: 10.1121/1.4744938

Gradshteyn, I. S., & Ryzhik, I. M. (2000). *Table of integrals and products* (6th Ed.). New York: Academic

Hasanpour Tadi, S., & Shokri, B. (2024). Perfect cylindrical cloak under gyration, non-inertial effects make perfect cloak visible. *Waves in Random and Complex Media*, *34*(1), 20-32. doi: 10.1080/17455030.2021.1898697

Huang, X., Zhong, S., & Stalnov, O. (2014). Analysis of scattering from an acoustic cloak in a moving fluid. *The Journal of the Acoustical Society of America*, *135*(5), 2571-2580. doi: 10.1121/1.4869815

Iemma, U. (2016). Theoretical and numerical modeling of acoustic metamaterials for aeroacoustic applications. *Aerospace, 3*(2), 15. doi: 10.3390/aerospace3020015

Iemma, U., & Palma, G. (2017). On the use of the analogue transformation acoustics in aeroacoustics. *Mathematical Problems in Engineering*, *2017*. doi: 10.1155/2017/8981731

Iemma, U., & Palma, G. (2018). Convective correction of metafluid devices based on Taylor transformation. *Journal of Sound and Vibration*, *443*, 238–252. doi: 10.1016/j.jsv.2018.11.047

Iemma, U., & Palma, G. (2020). Design of metacontinua in the aeroacoustic spacetime. *Scientific Reports*, *10*(1). doi: 10.1038/s41598-020-74304-5

Knobloch, K., Manoha, E., Atinault, O., Barrier, R., Polacsek, C., Lorteau, M., ... & Enghardt, L. (2022). Future aircraft and the future of aircraft noise. In *Aviation Noise Impact Management: Technologies, Regulations, and Societal Well-being in Europe* (pp. 117-139). Cham: Springer International Publishing.

MATLAB. (2024). *ode23s*. https://uk.mathworks.com/help/matlab/ref/ode23s.html

Morse, P. M., & Ingard, K. U. (1968). *Theoretical acoustics*. New Jersey: Princeton University Press.

Norris, A. N. (2008). Acoustic cloaking theory. *Proceedings of the Royal Society A: Mathematical, Physical and Engineering Sciences*, *464*(2097), 2411-2434. doi: 10.1098/rspa.2008.0076

Palma, G., Mao, H., Burghignoli, L., Göransson, P., & Iemma, U. (2018). Acoustic metamaterials in aeronautics. *Applied Sciences, 8*(6), 971. doi: 10.3390/app8060971

Palma, G., & Iemma, U. (2023). A metacontinuum model for phase gradient metasurfaces. *Scientific Reports*, *13*(1), 13038. doi: 10.1038/s41598-023-39956-z

Pendry, J. B., Schurig, D., & Smith, D. R. (2006). Controlling electromagnetic fields. *Science*, *312*(5781), 1780-1782. doi: 10.1126/science.1125907

Ruan, Z., Yan, M., Neff, C. W., & Qiu, M. (2007). Ideal cylindrical cloak: perfect but sensitive to tiny perturbations. *Physical Review Letters*, *99*(11), 113903. doi: 10.1103/PhysRevLett.99.113903

Sánchez-Dehesa, J., & Torrent, D. (2013). Transformation Acoustics. In Craster, R. V., & Guenneau, S. (Eds.), *Acoustic metamaterials*. (pp. 219-239). doi: 10.1007/978-94-007-4813-2_9

Torrent, D., & Sánchez-Dehesa, J. (2008). Acoustic cloaking in two dimensions: A feasible approach. *New Journal of Physics*, *10*(6), 063015. doi: 10.1088/1367-2630/10/6/063015

World Health Organization. (2022). *Guidance on environmental noise.* https://www.who.int/tools/compendium-on-health-and-environment/environmental-noise

Zhong, S., & Huang, X. (2013). Acoustic Cloaking in a Mean Flow. In *19th AIAA/CEAS Aeroacoustics Conference* (p. 2131).